\documentclass[traditabstract]{aa}  
\usepackage{graphicx}
\usepackage{txfonts}
\usepackage{rotating,tabularx,pdflscape}
\usepackage{supertabular}
\usepackage{caption}
\usepackage{natbib,twoopt}
\usepackage[breaklinks=true]{hyperref} 
\bibpunct{(}{)}{;}{a}{}{,} 
\newcommandtwoopt{\citeads}[3][][]{\href{http://adsabs.harvard.edu/abs/#3}%
  {\citealp[#1][#2]{#3}}}
\newcommandtwoopt{\citepads}[3][][]{\href{http://adsabs.harvard.edu/abs/#3}%
  {\citep[#1][#2]{#3}}}
\newcommandtwoopt{\citetads}[3][][]{\href{http://adsabs.harvard.edu/abs/#3}%
  {\citet[#1][#2]{#3}}} 
\newcommandtwoopt{\citeyearads}[3][][]%
                 {\href{http://adsabs.harvard.edu/abs/#3}{\citeyear[#1][#2]{#3}}} 

\def\kms    {\ifmmode{{\rm \ts km\ts s}^{-1}}\else{\ts km\ts s$^{-1}$}\fi}
\def\lsol   {\ifmmode{{\rm L}_{\odot}}\else{L$_{\odot}$}\fi}
\def\msol   {\ifmmode{{\rm M}_{\odot}}\else{M$_{\odot}$}\fi}
\def\hi     {\ifmmode{{\rm H}{\rm \small I}}\else{H\ts {\scriptsize I}}\fi}
\def\hh   {\ifmmode{{\rm H}_2}\else{H$_2$}\fi}

\def\zsol   {\ifmmode{{\rm Z}_{\odot}}\else{Z$_{\odot}$}\fi}
\def\tex {\ifmmode{{T}_{\rm ex}}\else{$T_{\rm ex}$}\fi}
\def\tmb {\ifmmode{{T}_{\rm mb}}\else{$T_{\rm mb}$}\fi}

\begin{document} 
\title{M31 circum-nuclear region: a molecular survey with the IRAM-interferometer}
\titlerunning{M31 circum-nuclear region}
\author{Julien Dassa-Terrier\inst{1}, Anne-Laure Melchior\inst{1}, Fran\c coise Combes\inst{1,2}}
\institute{Sorbonne Universit{\'e}, Observatoire de Paris, Universit{\'e} PSL, CNRS, LERMA, F-75014, Paris, France\\
  \email{Julien.Dassa-Terrier@obspm.fr,A.L.Melchior@obspm.fr,Francoise.Combes@obspm.fr}
  \and
 Coll\`ege de France, 11, Place Marcelin Berthelot, F-75\,005 Paris, France}\date{Received August 10, 2018; accepted February 13, 2019}
\abstract {We analyse molecular observations performed at IRAM interferometer in CO(1-0) of the circum-nuclear region (within ~250 pc) of Andromeda, with 2.9’’ = 11 pc resolution. We detect 12 molecular clumps in this region, corresponding to a total molecular mass of $(8.4\pm 0.4) \times 10^4\,M_\odot$. They follow the Larson's mass-size relation, but lie well above the velocity-size relation. We discuss that these clumps are probably not virialised, but transient agglomerations of smaller entities that might be virialised. Three of these clumps have been detected in CO(2-1) in a previous work, and we find temperature line ratio below 0.5. With a RADEX analysis, we show that this gas is in non local thermal equilibrium with a low excitation temperature ($T_{ex} = 5-9\,K$). We find a surface beam filling factor of order 5\,$\%$ and a gas density in the range $60-650$\,cm$^{-3}$, well below the critical density.  With a gas-to-stellar mass fraction of $4\times 10^{-4}$ and dust-to-gas ratio of 0.01, this quiescent region has exhausted his gas budget. Its spectral energy distribution is compatible with passive templates assembled from elliptical galaxies. While weak dust emission is present in the region, we show that no star formation is present and support the previous results that the dust is heated by the old and intermediate stellar population. We study that this region lies formally in the low-density part of the Kennicutt-Schmidt law, in a regime where the SFR estimators are not completely reliable. We confirm the quiescence of the inner part of this galaxy known to lie on the green valley.}


\maketitle

\section{Introduction} 
The evolution of the gas content and star formation activity in the central kiloparsec of galaxies is key to understand the coupling of the black hole evolution with the rest of the galaxy. Beside the activity of the central engine probably powered by mass accretion \citep[e.g.][]{1969Natur.223..690L,1995PASP..107..803U}, the properties of the host galaxy are directly impacted by this so-called AGN feedback \citep[][and references therein]{2012ARA&A..50..455F}. The scaling relations, between the supermassive black hole mass and the bulge mass and velocity dispersion of the host \citep[e.g.][]{2000ApJ...539L...9F,2002ApJ...574..740T,2003ApJ...589L..21M,2009ApJ...698..198G,2013ARA&A..51..511K,2013ApJ...764..184M}, suggest a close connection between supermassive black holes and their hosts \citep[e.g.][]{1998A&A...331L...1S,2005Natur.433..604D,2015ARA&A..53..115K}. The growth of supermassive black holes over cosmic time is probably dominated by external gas accretion \citep[e.g.][]{1982MNRAS.200..115S, 2006MNRAS.365...11C}. Indeed, galaxy mergers \citep[e.g.][]{1991ApJ...370L..65B,2005MNRAS.361..776S,2006ApJS..163....1H} or galactic bars \citep[e.g][]{1990ApJ...363..391P,2006MNRAS.370..289B,2010MNRAS.407.1529H}  may efficiently transport gas towards the galactic nucleus through gravitational torques \citep[e.g][]{2005A&A...441.1011G} over a few dynamical times. 

While the total AGN activity is correlated with the global star formation rate (SFR) as a function of cosmic time \citep[e.g.][]{2004ApJ...613..109H,2014ARA&A..52..589H}, SFR has been mostly quenched in the local Universe \citep[e.g.][]{2016MNRAS.461.3111B} while the activity of central black holes has been much reduced \citep[e.g.][]{2015MNRAS.452..575S}. The mechanisms responsible for this quenching are currently investigated. Relying on simulations, \citet{2017MNRAS.465...32B} show these red and blue sequences of galaxies result from a competition between star formation-driven outflows and gas accretion on to the supermassive black hole at the galaxy's centre. \citet{2016MNRAS.461.3111B} argue that the quenching occurred inside out, and that the star formation stopped first in the central region. However, these inside-out mechanisms are probably not universal as in dense environments, like groups or clusters of galaxies, the quenching could also occur outside-in. \citet{2015Natur.521..192P} argue that the strangulation mechanism in which the supply of cold gas to the galaxy is halted is the main mechanism responsible for quenching star formation in local galaxies with $M_* < 10^{11} M_\odot$. In the GASP (GAs Stripping Phenomena in galaxies with MUSE) survey, \citet{2017ApJ...844...48P} are exploring in the optical more than hundred galaxies in different environments, with the idea to get better constraints on both types of scenarios. \citet{2017ApJ...846...27G} have shown that gas is stripped out in JO204 a jellyfish galaxy in A957 and the star formation activity is reduced in the outer part. With a study based on local galaxies, \citet{2018arXiv180505352F} argue that AGN-driven outflows are likely capable of clearing and quenching the central region of galaxies. 

With a stellar mass of $5\times 10^{10} \, M_\odot$\citep{2014A&A...567A..71V}, Andromeda belongs to the transition regime between the active blue-sequence galaxies and passive red-sequence galaxies \citep[e.g.][]{2017MNRAS.465...32B,Baldry2006} which happens around the stellar mass of 3 $\times$ 10$^{10}$ M$_\odot$ \citep[e.g.][]{Kauffmann2003}. It is a prototype galaxy from the Local Group where the star formation has been quenched in the central part. It hosts both very little gas and very little star formation, while the black hole is basically quiet, with some murmurs \citep{2011ApJ...728L..10L}. In a previous study about M31 nucleus, \citet{Melchior:2017} show that there is no gas within the sphere of influence of the black hole. Indeed, the gas has been exhausted. Most scenarios of the past of evolution of Andromeda reproduce the large scale distribution, and show evidence of a rich collision past activity \citep{2001Natur.412...49I,2004ApJ...601L..39T,2006ApJ...638L..87G,2009Natur.461...66M,2014ApJ...780..128I,2016ApJ...827...82M,2018MNRAS.475.2754H}. However, the exact mechanism quenching the activity in the central kilo-parsec is still unknown \citep{2017A&A...600A..34T}. \citet{2006Natur.443..832B} proposed a frontal collision with M32, which could account for the 2 ring structures, observed in the dust distribution. \citet{2011A&A...536A..52M} and (\citeyear{2016A&A...585A..44M}) show the presence of gas along the minor axis and support the scenario of the superimposition of an inner 1-kpc ring with an inner disc. \citet{2013A&A...549A..27M} estimated a minimum total mass of $4.2 \times 10^4 M_\odot$ of molecular gas within a (projected) distance to the black hole of 100\,pc. This is several orders of magnitude smaller than the molecular gas present in the Central Molecular Zone of the Milky Way \citep{2000ApJ...545L.121P,2011ApJ...735L..33M}. In the Galaxy, while large amounts of dense gas are present in the central region, \citet{2014MNRAS.440.3370K} discuss the different processes that combine to inhibit the star formation, observed a factor 10 times weaker than expected \citep[e.g.][]{2008AJ....136.2782L}.

In this article, we analyse new molecular gas observations obtained with IRAM Plateau-de-Bure interferometer achieving a 11\,pc resolution. This resolution well below the typical size where a tight correlation is observed between star formation rate and gas density (K-S law). We study the properties of this gas and how it relates with the star formation activity in this region.

In Sect. \ref{sect:obs}, we present the analysis of the data cube, enabling an automatic selection of 12 molecular clouds. In Sect. \ref{sect:clprop}, we discuss the properties of these clumps. In Sect. \ref{sect:disc}, we discuss how this detection of molecular clouds in the equivalent of the Central Molecular Zone correlates with the information available on the star formation activity.

\section{Data analysis}
\label{sect:obs}
We here describe the identification of molecular clumps in the IRAM-PdB data cube. 
In Sect. \ref{ssect:obsdared}, we describe the IRAM-PdB observations and data reduction. In Sect. \ref{ssect:core}, we select the core of significant clumps above a given threshold and discuss their significance. In Sect.~\ref{ssect:3Dsize}, we compute the 3D size, velocity dispersion and total flux of these core clumps following the method proposed by \citet{Rosolowsky2006}. In Sect.~\ref{ssec:velo}, we discuss the velocity distribution of the detections and proceed to clean side lobes. In Section~\ref{ssect:PCA}, we apply the principal component analysis statistical method to discriminate reliable candidates using the descriptive parameters of the previous subsections. A final check is performed with a visual inspection of each selected clump. Last, in Sect.~\ref{ssec:FinalSelec}, we sum up our procedure and present our final selection.

\subsection{Observations and data reduction}
\label{ssect:obsdared}
We present here an analysis of an IRAM-PdB interferometer mosaic of 4 fields covering the centre of Andromeda and a field of view of about 2 arcmin off-centred to the South-East. As described in \citet{Melchior:2017}, the initial motivation of these observations was to search for molecular gas next to the black hole, and the region corresponding to the sphere of influence of M31's black hole, i.e. $R_{SOI}\sim 14$\,pc, has been explored. A 2\,mJy signal with a line-width of 1000\,km\,s$^{-1}$ was expected by \citet{2007ApJ...668..236C}, but were excluded at a 9\,$\sigma$ level. Only a small 2000\,$M_\odot$ clump,  lying most probably outside the sphere of influence of the black hole, has been detected, and is seen in projection. In this article, we extend this first exploration to the whole data cube that covers the equivalent of the Central Molecular Zone for the Milky Way \citep{1996ARA&A..34..645M, 1996ApJ...460..334O}. The observations have been performed in 2012 with the 5-antenna configuration with the WideX correlator. It covers a wide velocity band $[-3000,6000]$\,km\,s$^{-1}$, with a velocity resolution of $\delta v = 5.07$\,km\,s$^{-1}$ and a pixel size of 0.61$\arcsec$. Four fields have been observed at the positions provided in Table \ref{tab:fields} and integrated for about 4 hours. A standard reduction has been performed with the GILDAS software and these fields have been combined. We perform our analysis on the mosaic assembled with a beam of 3.37$\arcsec\,\times\,$2.45$\arcsec$ (PA $=70^{\circ}$) and a pixel size of 0.61$\arcsec$. We focus our work on the $[-600,0]$\,km\,s$^{-1}$ velocity band, binned in 119 channels. 

\begin{table*}
 \caption{Positions of the 4 fields observed at IRAM-PdB in 2012. The phase centre of the observation was 00$^h$42$^m$44.99$^s$,\,$+$41$^\circ$15$\arcmin$57.8$\arcsec$. Beside the J2000.0 equatorial coordinates of the centre of each field, we also provide the number of visibilities acquired per field, and the corresponding beam and pixel size (square). The sampling of the 4th field is worse than for the other ones, which explains a larger level of noise.}                     
  \label{tab:fields}        
  \centering                
  \begin{tabular}{rrrrrr}   
    \hline\hline            
Field & RA & DEC & Visibilities & Beam & Pixels\\
\hline\hline
1 & 00$^h$42$^m$44.39$^s$ & $+$41$^\circ$16$^\prime$08.3$^{\prime\prime}$ & 4473 & 3.37$^{\prime\prime}\times$2.45$^{\prime\prime}$ &0.61$^{\prime\prime}\times$0.61$^{\prime\prime}$\\
2 & 00$^h$42$^m$46.78$^s$ & $+$41$^\circ$16$^\prime$12.3$^{\prime\prime}$ & 4432 &3.37$^{\prime\prime}\times$2.45$^{\prime\prime}$&0.61$^{\prime\prime}\times$0.61$^{\prime\prime}$\\
3 & 00$^h$42$^m$45.27$^s$ & $+$41$^\circ$15$^\prime$54.3$^{\prime\prime}$ & 4743 &4.53$^{\prime\prime}\times$2.76$^{\prime\prime}$&0.74$^{\prime\prime}\times$0.74$^{\prime\prime}$\\
4 & 00$^h$42$^m$43.50$^s$ & $+$41$^\circ$15$^\prime$36.3$^{\prime\prime}$ & 3131 &4.53$^{\prime\prime}\times$2.76$^{\prime\prime}$&0.74$^{\prime\prime}\times$0.74$^{\prime\prime}$\\
\hline\hline
\end{tabular}
\end{table*}

\begin{table*}
 \caption{Summary of the identification of cores process. We select spectra with at least 2 consecutive channels deviating from 3$\sigma$ from the continuum level. We keep positive and negative signals at each stage, in order to have an estimate of the noise level. We provide the number of pixels and the number of core clumps derived. Only core clumps with at least 3 adjacent pixels are kept for subsequent analysis. The approximated mean flux is the mean value of the sums of detected pixels fluxes of all core clumps.}
  \label{tab:pixels_info}    
  \centering                 
  \begin{tabular}{lcccccccccc} 
    \hline\hline             
                & \multicolumn{3}{c}{Number of pixels}    & \multicolumn{3}{c}{Core clumps} & \multicolumn{2}{c}{Mean Surface (pixels)} & \multicolumn{2}{c}{Approx. mean flux (mJy/beam)} \\
                & Total & Positive & Negative             & Total & Positive & Negative     & Positive & Negative                       & Positive & Negative \\
    \hline\hline
    $<=$ 2 pixels & 142   & 65       & 77                   & 102   & 47       & 55           &          & & &                               \\
    $>$  2 pixels & 903   & 598      & 305                  & 101   & 54       & 47           & 11.1     & 5.6 & 766 & -313                  \\
    \hline\hline
 \end{tabular}
\end{table*}

\begin{figure}[h!]
    \centering
    \includegraphics[width=0.48\textwidth,clip]{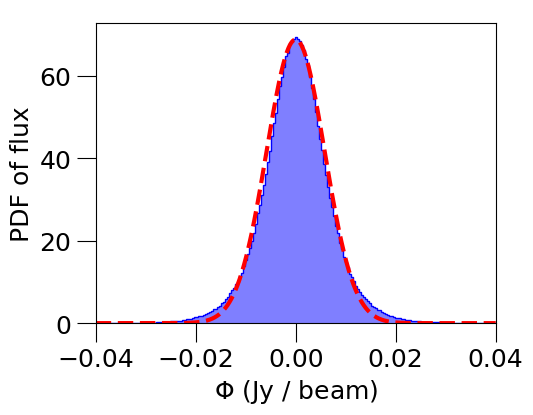}
    \includegraphics[width=0.48\textwidth,clip]{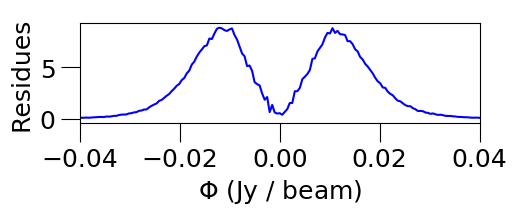}
    \caption{Up : Probability density function of the distribution of fluxes within the data cube and the associated Gaussian fit (red dashed curve). $\phi$ corresponds to the flux of each pixel (in the 3D datacube), and the spatial variation of the noise has not been corrected at this stage.  Down : residues associated to the distribution and its Gaussian fit. We observe that the distribution is not Gaussian, something we expect from a cube containing signal. The residues are mostly symmetrical, showing that the signal is very faint and needs advanced detection methods.}
    \label{fig:FluxDistrib}
\end{figure}

A quick view of the data cube confirms the absence of large amount of gas in this region. However, there are numerous clumps of molecular gas with a low velocity dispersion. In Figure~\ref{fig:FluxDistrib}, we observe a non-Gaussian flux distribution typical of a data cube with signal. We optimise a selection procedure to disentangle genuine clumps from noise. Given this configuration, we perform a 1-iteration CLEAN procedure to correct for the primary beam. We do not have single dish observations in CO(1-0) of this region. So we cannot correct from short spacing. In addition, as the signals are not extended and relatively weak, it is not possible to analyse the data cube with standard algorithms, e.g. the signal characteristics do not match the criteria stated in \citet{Rosolowsky2006}. Hence, we perform a basic signal detection in the data cube in an automatic way, and check our results with careful visual inspections and classical methods (e.g. GILDAS). 

\subsection{Detection of 3$\sigma$ peaks}
\label{ssect:core}
We rely on the first and second moments of each spectra with a $3\sigma$ clip to get a best estimate of the mean flux $\langle\Phi\rangle{(x,y)}$ and rms noise level $\sigma_\text{noise} \left(x,y\right)$, where $(x,y)$ refers to the pixel position. We will refer to these parameters simply as $\langle\Phi\rangle$ and $\sigma_\text{noise}$ hereafter. The noise level, which map is displayed in \citet{Melchior:2017}, is at its lowest close to the black hole about 3.2\,$mJy/beam$ at the velocity resolution (5.07\,km/s), and increases towards the edges of the field of view (up to 13\,$mJy/beam$).
In Sect. \ref{sect:cont},  we derive 3$\sigma$ upper limit on the continuum level: about 15 $\mu$Jy within 15$\arcsec$ estimated on the whole bandwidth $[-3000,6000]$\,km\,s$^{-1}$.
\begin{figure}[h!]
    \centering
    \includegraphics[width=0.48\textwidth,clip]{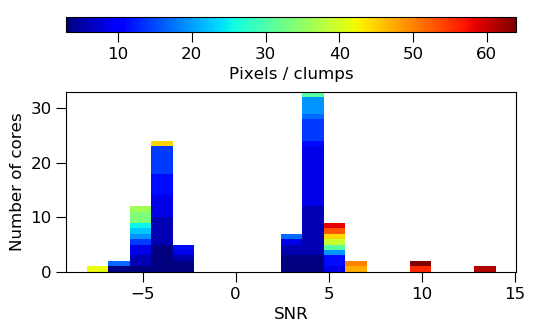}
    \caption{ Distribution of the S/N ratio based on the 3$\sigma$ peak flux ($SNR_\text{peak}$) for positive and negative 3-$\sigma$ clumps. Each clump is associated with a colour based on the number of pixels detected in the core. While higher $SNR_\text{peak}$ are observed for positive detections, we still have a noise contribution affecting both the positive and negative sides.}
    \label{fig:histoSNR}
\end{figure}

We select pixels which spectrum exhibits more than 2 consecutive spectral channels with a flux excess or deficit of 3\,$\sigma_\text{noise}$. At this stage, we keep the negative signals in order to access the coherence of the detections with respect to noise. As detailed in Table~\ref{tab:pixels_info}, we detect 1045 pixels: 663 (resp. 382) with a positive (resp. negative) signal. As isolated pixels or aggregates of two pixels are most likely noise, we define clump candidates as groups of at least 3 adjacent pixels exhibiting more than 2 consecutive channels at 3\,$\sigma_\textrm{noise}$ from the mean level. In principle, this criteria selects  the core of possible clumps, i.e. the central 3\,$\sigma$ pixels with high S/N ratio. 142 pixels (65 positive and 77 negative) with less than 3 adjacent pixels are excluded. We thus keep 598 (resp. 305) pixels gathered in 54 positive (resp. 47 negative) 3-$\sigma$ clumps. 
33.8\% of the detected signals are negative. Positive (resp. negative) 3-$\sigma$ clumps have a mean flux of 766\,mJy/beam (resp. -313\,mJy/beam). Their average surface is of 11.1 pixels (to be compared to 6.5 for the negative 3-$\sigma$ clumps). The number of pixels at half-power beam width (HPBW) is about 25 pixels, hence we only see the inner parts of the clumps. Given the relatively small number of pixels in the 3-$\sigma$ clumps compared to 25, we expect that the detections are close to the spatial resolution. Figure~\ref{fig:veloStudy} displays the mean velocity map of positive (left panel) and negative (right panel) 3-$\sigma$ peak detections above 3$\sigma_\textrm{noise}$.

We estimate the peak S/N ratio (SNR$_\textrm{peak}$) associated to detected 3-$\sigma$ clumps as the ratio of the peak flux $\Phi_\textrm{peak}$ of the 3-$\sigma$ clump to the noise estimated for its spectra. We show this S/N distribution in Figure~\ref{fig:histoSNR}. The average SNR$_\textrm{peak}$ for the positive (resp. negative) 3-$\sigma$  clumps is 4.4 (resp. -3.7) with 28 positive (resp. 16 negative) 3-$\sigma$ clumps showing a peak signal higher than 4\,$\sigma_\textrm{noise}$ (resp. lower than -4\,$\sigma_\textrm{noise}$), and 3 positive 3-$\sigma$  clumps with peak S/N larger than 8. The clumps we detect are thus significantly brighter than the level of noise, estimated with the negative clumps. Although the two sets behave differently, more investigations are required to select with confidence genuine molecular clouds. In the three next subsections, we calculate global properties, exclude side lobes and apply principal component analysis to make our final selection.

\subsection{Physical quantities derived from moment measurements}
\label{ssect:3Dsize}
\begin{figure*}[h!]
    \centering
    \includegraphics[width=0.96\textwidth,clip]{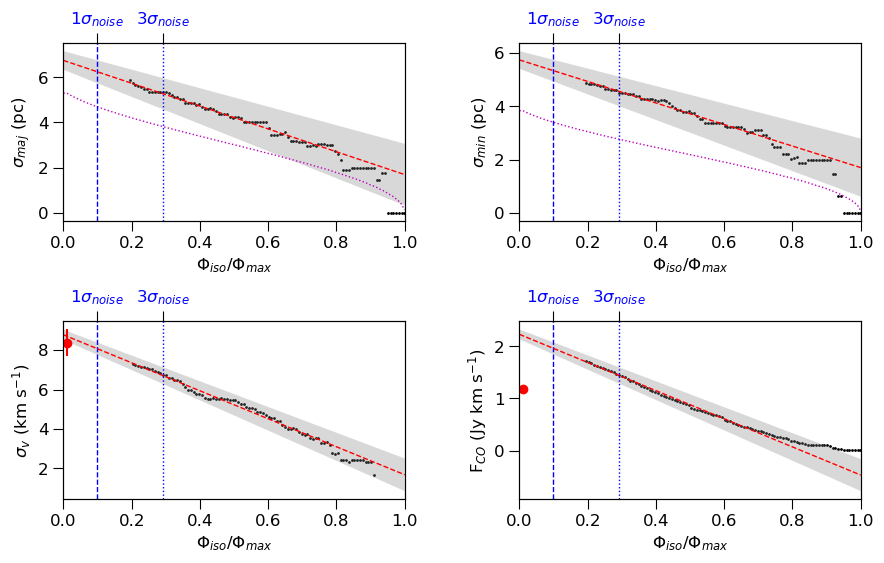}
    \caption{Extrapolation method applied on the highest S/N clump at J2000 coordinates $00^h\,42^m\,43.9^s$ and $+41^{\circ}\,15'\,35.84^{\prime\prime}$ with SNR$_\text{peak}=10.3$ and SNR$_\text{tot}=64.4$.
    We perform extrapolations to the $0-$level with a linear regression (red dashed line) on the data for spatial size $\sigma_\text{maj}$ and $\sigma_\text{min}$ (top panels), velocity dispersion $\sigma_\text{v}$ distribution (bottom left panel) and total flux $F_\text{CO}$  (bottom right panel). The uncertainty is computed with the bootstrapping method. With blue dashed lines we indicate where $\Phi_\text{iso} / \Phi_\text{max}$ is equal to 1 and 3$\sigma_\text{noise}$. Since we expect our clumps to be barely resolved, we show as a magenta dashed line the size for a perfect Gaussian PSF, noiseless, simulated equivalent case. A red dot indicates, for the velocity distribution and the total flux, the values found by fitting a Gaussian in GILDAS. We observe that the size over the major and minor axis is slightly larger than the one obtained for the simulated clump, which suggests that this clump is barely resolved. The velocity dispersion is about 4 times the spectral resolution (2.2\,km\,s$^{-1}$). There is a good agreement between the extrapolated and the one measured through GILDAS velocity dispersions while the total flux is slightly higher for the extrapolation.}
    \label{fig:Extrap1}
\end{figure*}
\begin{figure}[h!]
\centering
\includegraphics[width=0.48\textwidth,clip]{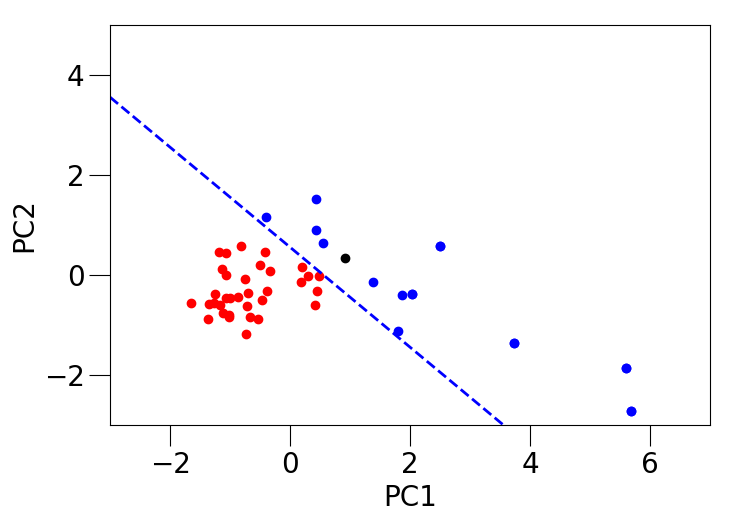}
\includegraphics[width=0.48\textwidth,clip]{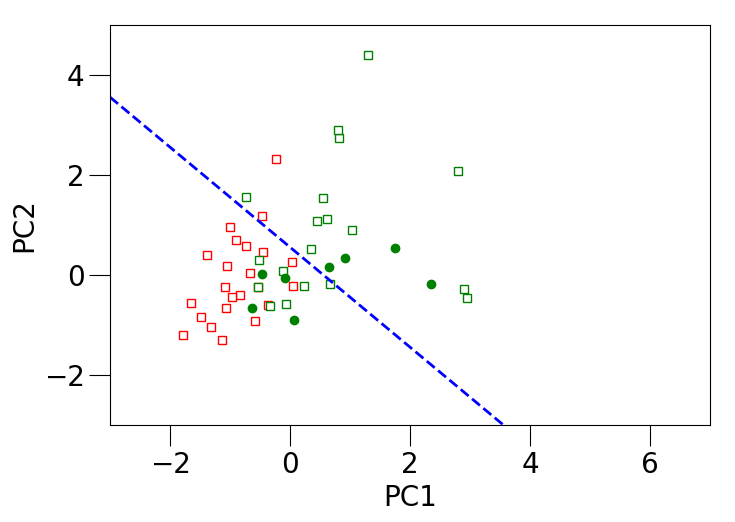}
\caption{Results of the PCA analysis. The principal components PC1 and PC2 are defined in Eq. \ref{eq:PCA}. Clumps with positive (resp. negative) signals are shown in full circles (resp. empty squares). {\em Top panel: } Map from the PCA for positive signals.  Selected (resp. rejected) clumps are in blue (resp. red).  The blue dashed straight line shows the limit between selected and rejected clumps and follows the relation: $PC2 = -1 \, \times \, PC1 \, + \, 0.55$.
{\em{Bottom panel:}} Map for rejected signals: negative signals and side lobes. Side lobes, identified in Sect. \ref{ssec:velo}, are displayed in green, and the other negative signals not identified as side lobes are displayed in red.   We can see that side lobes can have a high SNR. Most side lobes correspond to negative signals, but we also eliminate four side lobes with a positive signal. }
\label{fig:PCA}
\end{figure}
\begin{figure}[h!]
\centering
\includegraphics[width=0.48\textwidth,clip]{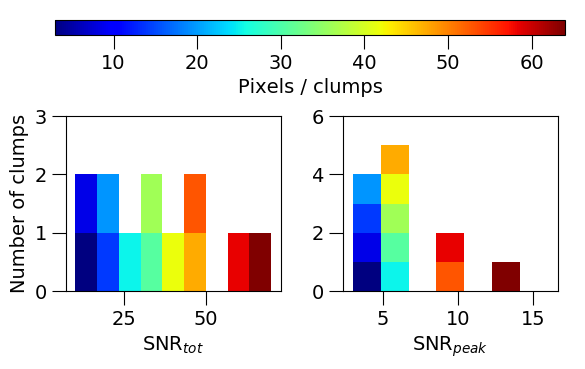}      
\caption{SNR distribution of the selected clumps. The distribution of $SNR_\text{tot}$ shows a clear correlation between cores with large number of pixels and high S/N ratio.}
\label{fig:snrtot}
\end{figure}

In order to further characterise the detection of molecular clouds, we estimate the clump sizes, velocity dispersions and total fluxes. \citet{Rosolowsky2006} note a strong correlation between the resolution and the measured size of clumps for those with a size close to the spatial resolution. This is probably the case here: most detected clumps are unresolved or close to the detection limit. We expect them to have an elliptical shape similar to the beam. Hence, we adapt the CPROPS (Cloud PROPertieS) method proposed by \citet{Rosolowsky2006} in order to optimise these measurements. We can note that \citet{Rosolowsky2006} have shown that this extrapolating method is optimum for a peak S/N larger than 10, while our data host only 2 clumps as strong as this. However, these authors have shown that the properties derived from interferometric data are underestimated with respect to single dish data, with a lost of order of 50\% in the integrated luminosity as already discussed by \citet{2000immm.proc...37S}.
\begin{figure*}
\centering
\includegraphics[width=0.33\textwidth,clip]{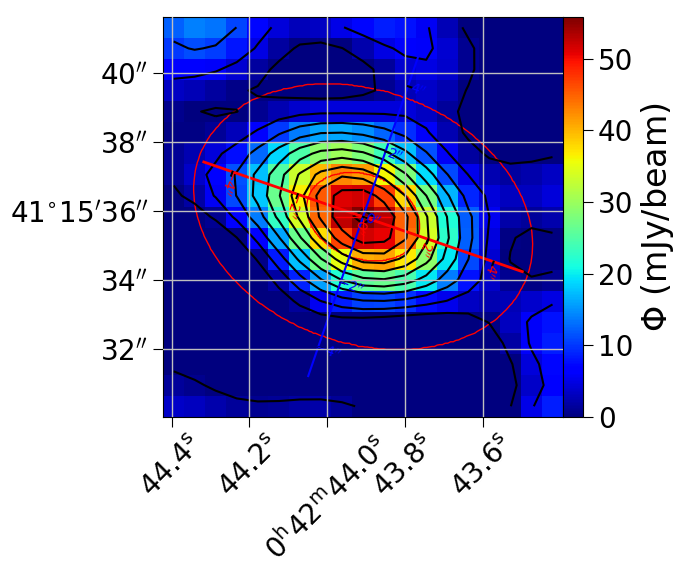}
\includegraphics[width=0.33\textwidth,clip]{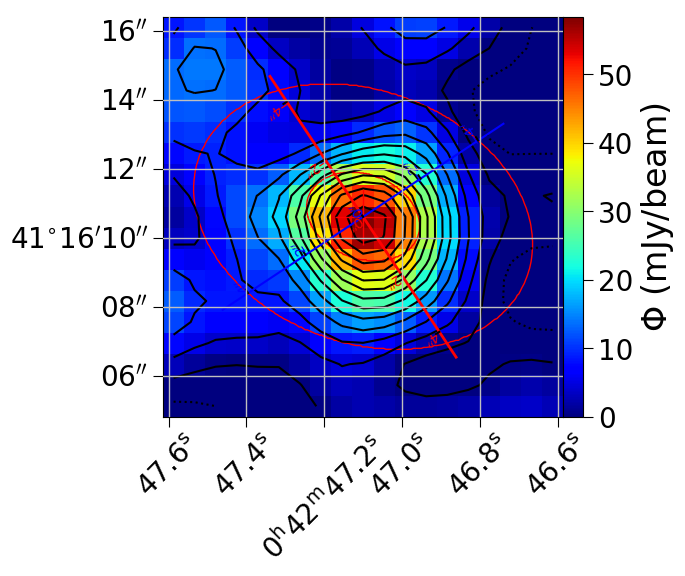}
\includegraphics[width=0.33\textwidth,clip]{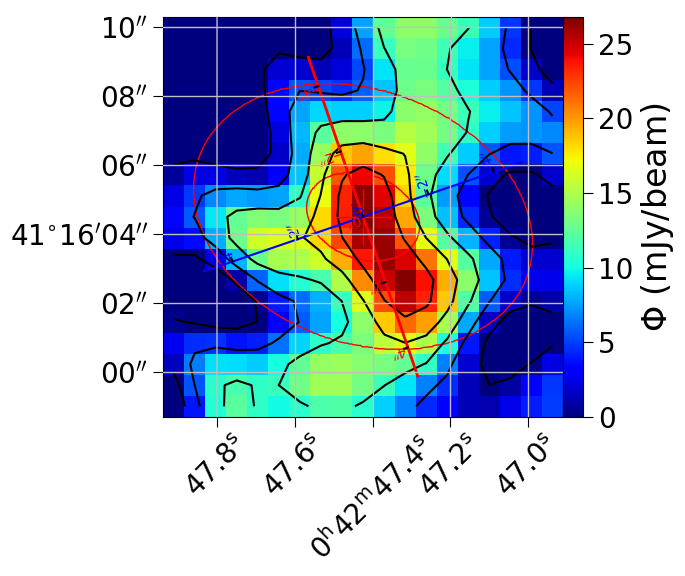}
\includegraphics[width=0.33\textwidth,clip]{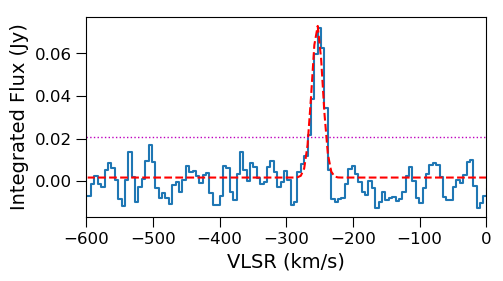}
\includegraphics[width=0.33\textwidth,clip]{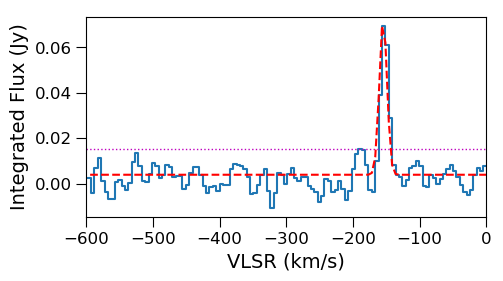}
\includegraphics[width=0.33\textwidth,clip]{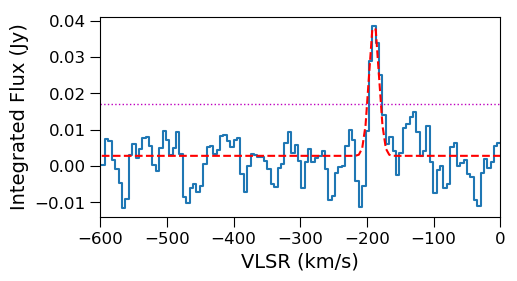} 
\includegraphics[width=0.33\textwidth,clip]{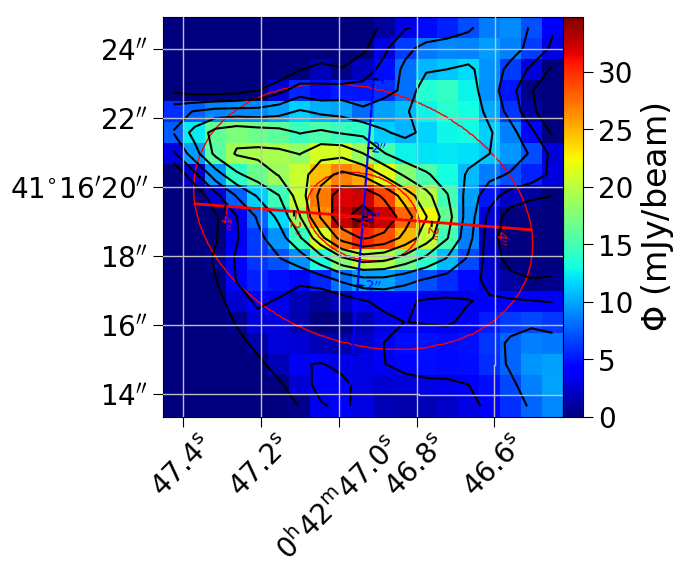}     
\includegraphics[width=0.33\textwidth,clip]{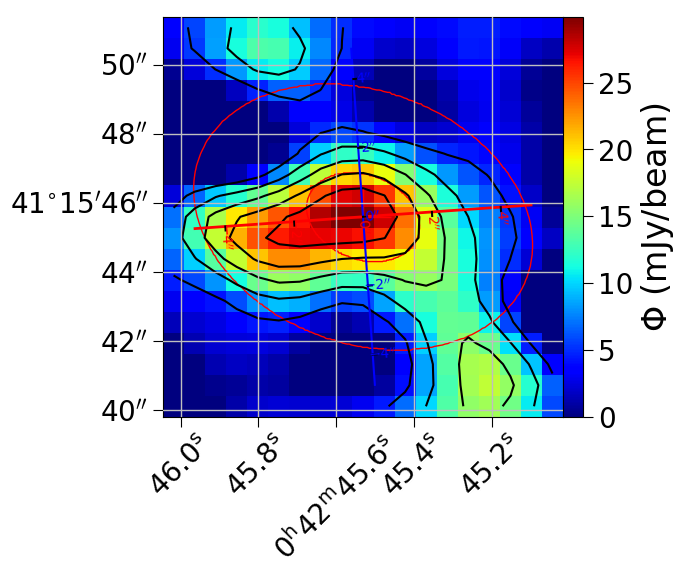}     
\includegraphics[width=0.33\textwidth,clip]{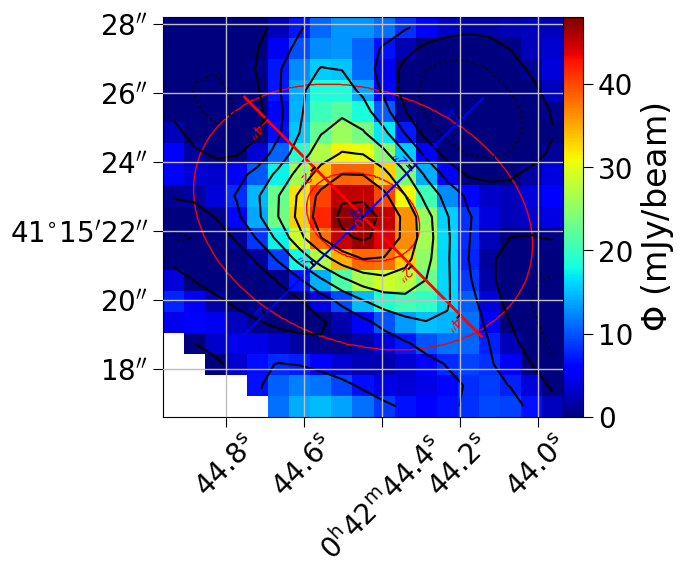}
\includegraphics[width=0.33\textwidth,clip]{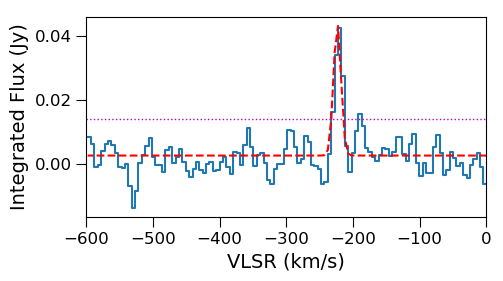}
\includegraphics[width=0.33\textwidth,clip]{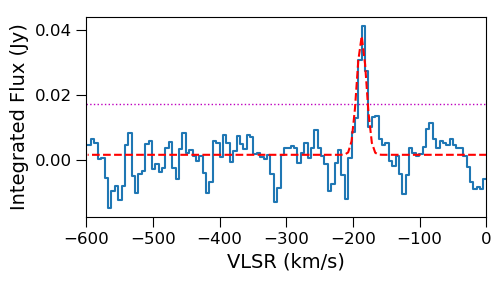}
\includegraphics[width=0.33\textwidth,clip]{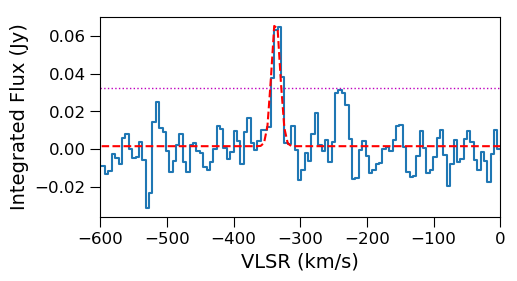} 
\includegraphics[width=0.33\textwidth,clip]{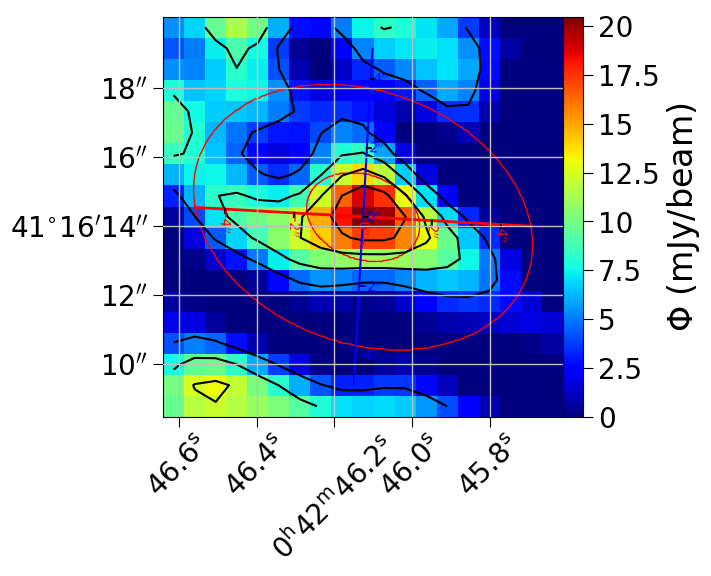}     
\includegraphics[width=0.33\textwidth,clip]{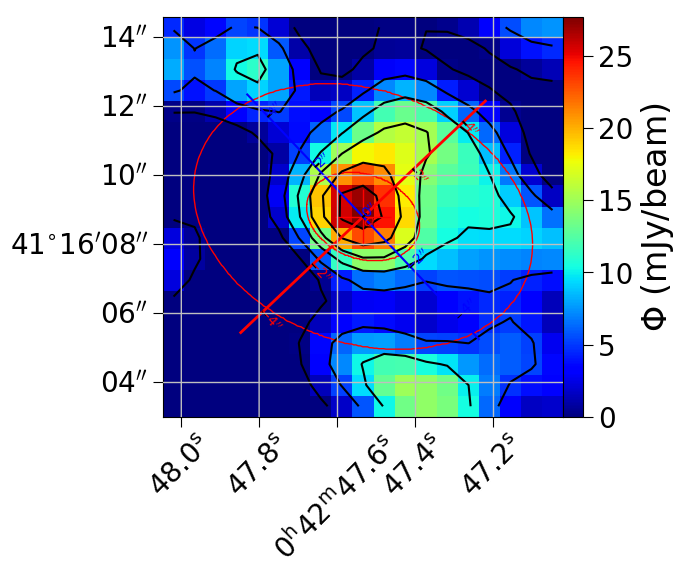}     
\includegraphics[width=0.33\textwidth,clip]{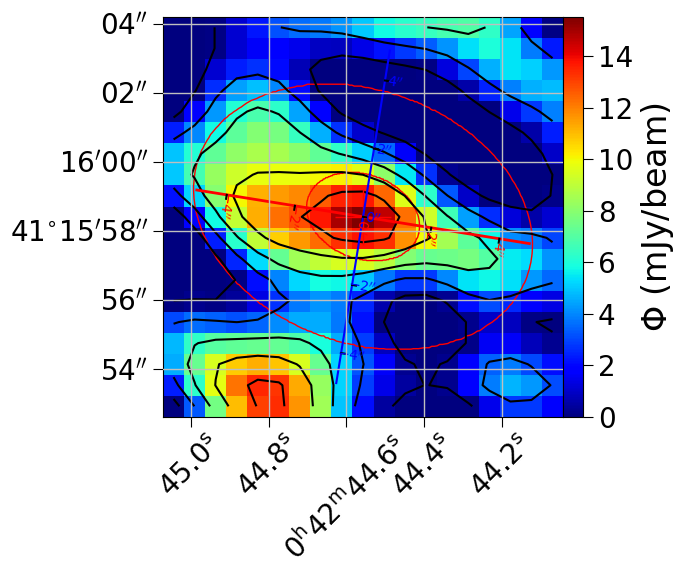}
\includegraphics[width=0.33\textwidth,clip]{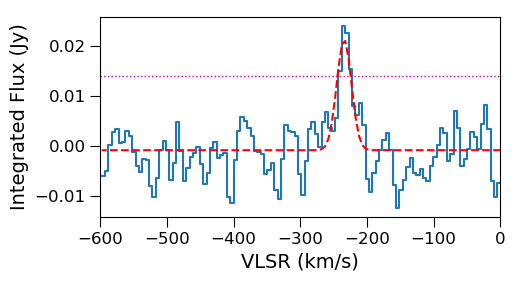}
\includegraphics[width=0.33\textwidth,clip]{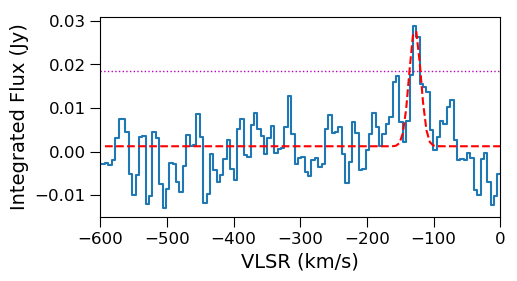}
\includegraphics[width=0.33\textwidth,clip]{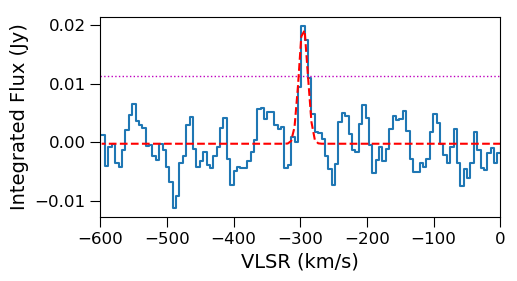} 
\caption{}
\end{figure*}
\begin{figure*}
\ContinuedFloat
\includegraphics[width=0.33\textwidth,clip]{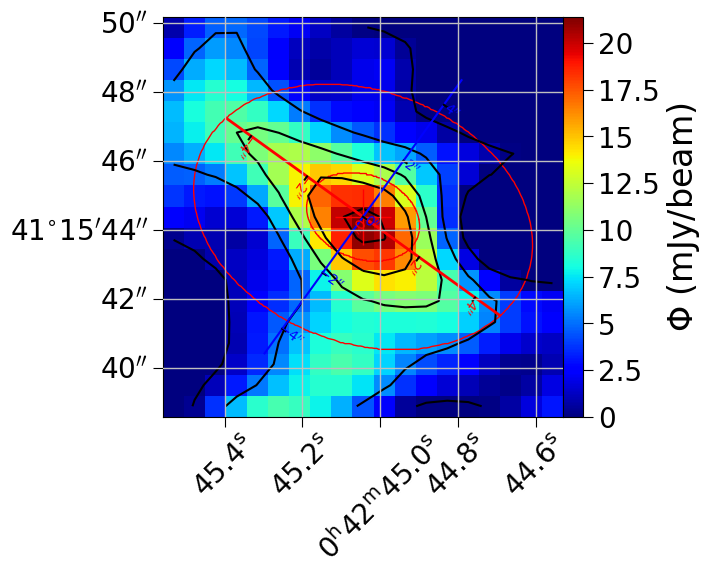}     
\includegraphics[width=0.33\textwidth,clip]{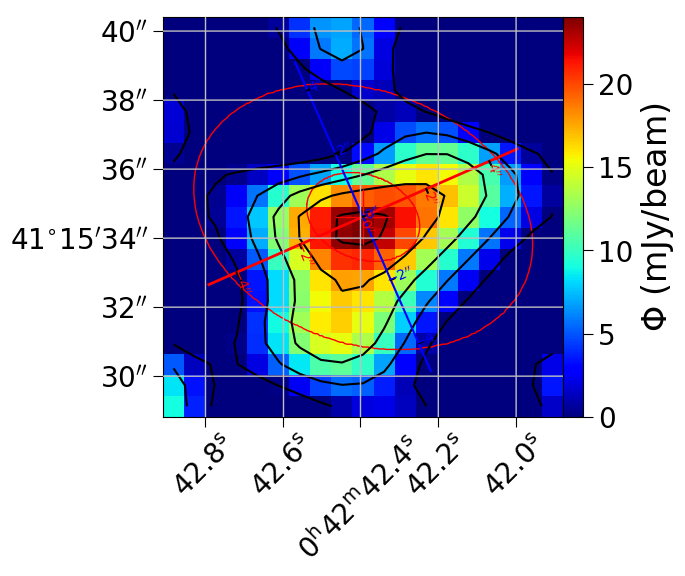}     
\includegraphics[width=0.33\textwidth,clip]{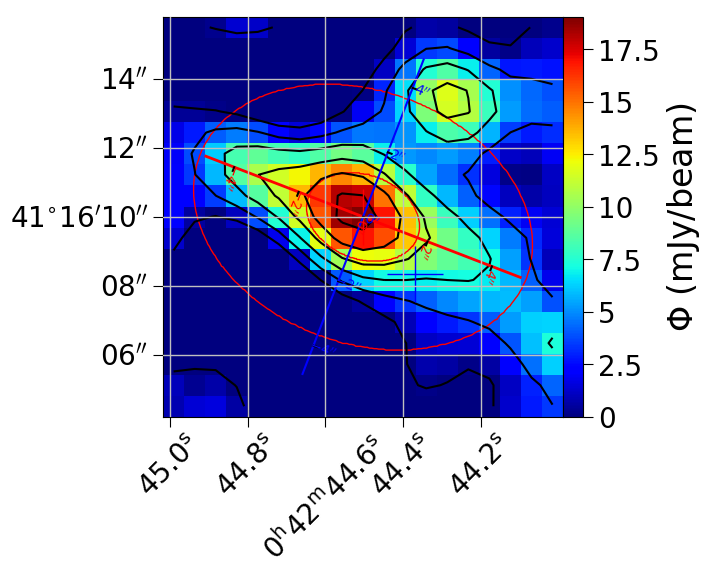}
\includegraphics[width=0.33\textwidth,clip]{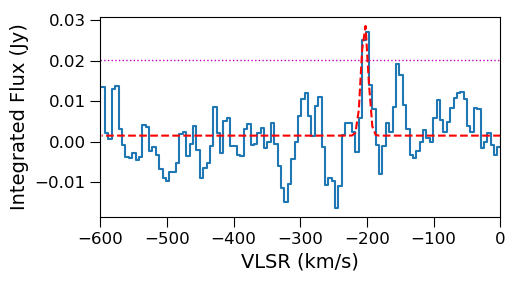}
\includegraphics[width=0.33\textwidth,clip]{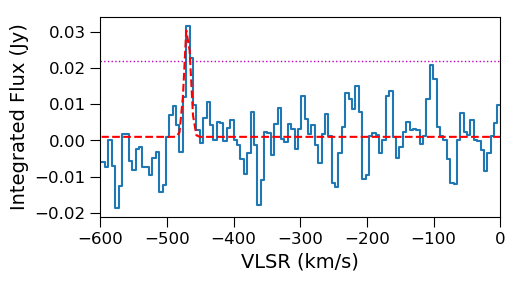}
\includegraphics[width=0.33\textwidth,clip]{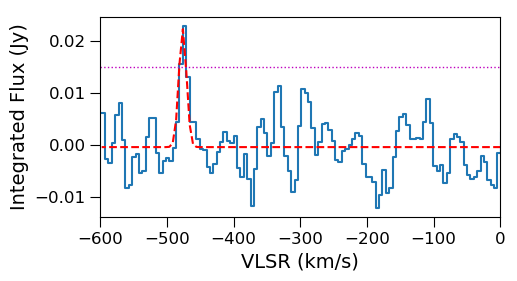}
\caption{Maps for the 12 clumps centred on the pixel and channel with peak flux and their associated integrated spectrum, organized by order of decreasing $SNR_\text{tot}$ (i.e. ID1 to ID12). The major and minor axes are displayed by a red and blue line respectively. Two ellipses show the area of the FWHM beam and of the box used for the calculation of extrapolation (with major and minor axis three times as big as the beam). The contours are separated by one sigma each, the negative sigma being show with dashed contours. \\
The spectra are displayed in blue. The magenta dashed line in the spectrum is the 3$\sigma$ level used for the selection procedure and the dashed red line corresponds to the Gaussian fit performed over the peak emission.}
\label{fig:Maps}
\end{figure*}

We calculate the second moments of the major x-axis and minor y-axis (defined for each clump). A principal component analysis, also performed with similar intent in \citet{Koda2006ApJ}, is used to define these axes: x and y refer to the major and minor axes respectively hereafter. For each clump, we define an elliptical box with major and minor axes twice as big as the beam centred on the pixel with peak flux $\Phi_\text{max}$. This intends to exclude surrounding regions with potential noise. We define a variable $\Phi_\text{iso}$, which corresponds to the equal-intensity isophotes. We, then, study the evolution of the size, the velocity dispersion, and total flux as a function of the intensity values corresponding to each isophote, varying from $\Phi_\text{max}$ to 2$\sigma_\text{noise}$. Up to 100 isophotes are thus used for each clump. We choose 100 isophotes since our box size may contain up to 60 pixels with 3 to 6 channels. Each value of $\Phi_\textrm{iso}$ delimits a region over which we sum the following moments over all pixels $i$ and channels $k$, with the major (resp. minor) axis size $\sigma_\text{maj}$ (resp. $\sigma_\text{min}$):
  \begin{equation}
  \sigma_\text{maj}\left(\Phi_\text{iso}\right)=\sqrt{\frac{\sum_{k}^{iso}\sum_{i}^{iso}\Phi_\text{i,k}[x_\text{i}-\overline{x}\left(\Phi_\text{iso}\right)]^2}{\sum_{k}^{iso}\sum_{i}^{iso}\Phi_\text{i,k}}},
  \end{equation}
  \begin{equation}
  \sigma_\text{min}\left(\Phi_\text{iso}\right)=\sqrt{\frac{\sum_{k}^{iso}\sum_{i}^{iso}\Phi_\text{i,k}[y_\text{i}-\overline{y}\left(\Phi_\text{iso}\right)]^2}{\sum_{k}^{iso}\sum_{i}^{iso}\Phi_\text{i,k}}},
  \end{equation}
where $\Phi_\text{i,k}$ is the flux in $Jy\,beam^{-1}$ over all channel $k$ and pixel $i$ respectively. $x_i$ and $y_i$ refer to the offset positions along the major and minor axis respectively. $\overline{x}$ and $\overline{y}$ are  the coordinate of barycentre of each isophote:
  \begin{equation}
  \overline{x}\left(\Phi_\text{iso}\right) = \frac{\sum_{k}^{iso}\sum_{i}^{iso}\Phi_\text{i,k} x_\text{i}}{\sum_{k}^{iso}\sum_{i}^{iso}\Phi_\text{i,k}},
  \overline{y}\left(\Phi_\text{iso}\right) = \frac{\sum_{k}^{iso}\sum_{i}^{iso}\Phi_\text{i,k} y_\text{i}}{\sum_{k}^{iso}\sum_{i}^{iso}\Phi_\text{i,k}}.
  \end{equation}

Similarly, the velocity dispersion can be defined as:
  \begin{equation}
  \sigma _\text{v}\left(\Phi_\text{iso}\right)=\sqrt{\frac{\sum_{k}^{iso}\sum_{i}^{iso}\Phi_\text{i,k}[v_\text{k}-\overline{v}\left(\Phi_\text{iso}\right)]^2}{\sum_{k}^{iso}\sum_{i}^{iso}\Phi_\text{i,k}}},
  \end{equation}
where :
  \begin{equation}
  \overline{v}\left(\Phi_\text{iso}\right) = \frac{\sum_{k}^{iso}\sum_{i}^{iso}\Phi_\text{i,k} v_\text{k}}{\sum_{k}^{iso}\sum_{i}^{iso}\Phi_\text{i,k}}
  \end{equation}
  and $v_\text{k}$ is the velocity at channel $k$. Last, the total integrated flux in $Jy\,km\,s^{-1}$ can be computed as:
  \begin{equation}
  F_\text{CO}\left(\Phi_\text{iso}\right)=\sum_{k}^{iso}\sum_{i}^{iso} \Phi_\text{i,k} \, \delta x \, \delta y \, \delta v,
  \end{equation}
where $\delta x$ and $\delta y$ are the size of the spatial pixels in arcsec and $\delta v$ is the spectral resolution in $km\,s^{-1}$. In order to find the values of the size, velocity dispersion and flux of the clumps, we use a weighted (by the number of pixels in each isophote) linear least-square regression to extrapolate the moments at $\Phi_\text{iso} = \Phi_0 = 0$ Jy/beam.  Figure~\ref{fig:Extrap1} (resp. Figure~\ref{fig:Extrap2}) displays the principle of this procedure applied to the determination of the parameters of the strongest cloud (resp. the cloud analysed in \citet{Melchior:2017}), we obtain the values of $\sigma_\text{maj}\left(\Phi_{0}\right)$, $\sigma_\text{min}\left(\Phi_{0}\right)$, $\sigma_\text{v}\left(\Phi_{0}\right)$ and F$_\text{CO}\left(\Phi_{0}\right)$. For comparison purposes, we also include values of the velocity dispersion and flux measured through Gaussian fit within the CLASS method in the software GILDAS. We do observe a good agreement between the two methods. In order to get consistent measurements we will subsequently use the extrapolated $\sigma_\text{v}$. \citet{Gratier2012} and \citet{Corbelli2017AA} adopt the \textsc{Class/Gildas} measurements for the velocity dispersions and integrated fluxes as they provide lower uncertainties. As further explained below, we do compute errors on the extrapolated values the same way as \citet{Rosolowsky2006}, with the bootstrap method and we do find uncertainties compatible with the \textsc{Class} measurements based on a Gaussian fit on the integrated spectra. 

The uncertainty over the extrapolated parameters is found through the bootstrapping method, which is a robust technique to estimate the error when it is difficult to formally determine the contribution of each source of noise. We proceed as follows: each isophote contains a number N of pixels, we choose randomly N pixels with replacement and so on for each isophote. A new extrapolation is made from this bootstrapped clump. This process is repeated 500 times and the standard deviation of the distribution of extrapolated values multiplied by the square root of the number of pixels in a beam, the oversampling rate (in order to account for pixels being correlated), is used as the uncertainty. The resulting uncertainties can be seen in Figure~\ref{fig:Extrap1} and in Table~\ref{tab:AllDat} which lists our final sample.

The root-mean-squared spatial size $\sigma_\text{r}$ of each clump is then calculated by deconvolving the spatial beam from second moments of the extrapolated clump size, assuming the clumps have the shape of a 2D Gaussian, and substracting the RMS beam sizes $\sigma_\text{beam,maj}$ and $\sigma_\text{beam,min}$:
  \begin{equation}
  \sigma_\text{r} = \left( \sqrt{\sigma_\text{maj}^2\left(\Phi_{0}\right) - \sigma_\text{beam,maj}^2} \sqrt{\sigma_\text{min}^2\left(\Phi_{0}\right) - \sigma_\text{beam,min}^2} \right)^{1/2},
  \end{equation}
  where $\sigma_\text{maj}\left(\Phi_{0}\right) > \sigma_\text{beam,maj}$ and $\sigma_\text{min}\left(\Phi_{0}\right) > \sigma_\text{beam,min}$, a condition met for all our sample, including negative clumps. Note that we compare the moments of major and minor size extrapolated at $\Phi_0$ to the RMS size of the beam, not to the projection at $\Phi_0$ of the beam size. This is a robust approach to take into account the resolution bias when clumps are barely resolved. This method shows high performances on marginal resolution, low S/N molecular clouds \citep{Rosolowsky2006}. While the beam has a position angle of 70$^{\circ}$, the mean inclination of the major axis of our clumps is (82$\pm$25)$^{\circ}$ indicating that some clumps are resolved and exhibit structures, most structures are unresolved. This will influence $\sigma_\text{r}$ calculation, hence we assume the clumps to be spherical which might lead to a slight underestimation of the spatial size.
  
  The equivalent radius R$_\text{eq}$ of the assumed spherical clump is chosen following the definition from \citet{1987ApJ...319..730S} : $R_{eq} \approx 1.91\sigma_r$. It was derived from the assumption that clumps behave as spherical clouds with a density profile $\rho \propto r^{-\beta}$ where $\sigma_r^2\left(\beta\right) = \frac{3 - \beta}{3(5 - \beta)} R^2$ and $\sigma_r^2\left(\beta = 1\right) = \sqrt{6} R^2\sim 2.45 R^2$ . The slightly different value found empirically by \cite{1987ApJ...319..730S} can be explained by the behaviour of the $^{12}$CO which displays a shallower density profile than the used model. As it is also used in \citet{Rosolowsky2006}, it provides an adequate approximation that permits comparison. From now on, we will refer to $\sigma_\text{maj}\left(\Phi_{0}\right)$, $\sigma_\text{min}\left(\Phi_{0}\right)$, $\sigma_\text{v}\left(\Phi_{0}\right)$ and F$_\text{CO}\left(\Phi_{0}\right)$, as $\sigma_\text{maj}$, $\sigma_\text{min}$, $\sigma_\text{v}$ and F$_\text{CO}$ since only the extrapolated values hold any relevance in this work.

From these properties, we can compute the S/N ratio $SNR_\text{tot}$ based on the total flux $F_\text{CO}$ and the noise of the spectrum integrated within the box containing each clump.

\subsection{Velocity coherence and side lobes}
\label{ssec:velo}
It is unlikely that two genuine clumps would have identical velocities, and we expect a velocity gradient up to 600\,km\,s$^{-1}$ in this region. Hence, it is highly probable that two clumps with the exact same velocity but different intensities correspond to one genuine clump and a side lobe which should be rejected from our selection. As seen in Figure~\ref{fig:veloStudy}, which shows the maps of positive and negative signals as well as the corresponding velocity distributions, a noticeable amount of clumps display similar or identical velocities.  The cleaning of side lobes is done through the \textsc{Mapping} method in the software \textsc{Gildas}. A \textsc{Clean} algorithm \citep{1995yera.conf....8G} is run over each channel where more than one clump is detected, with a support centred on the strongest clump. To assess the robustness of signals, we keep negative clumps and side lobes in a first step in order to explore their behaviour statistically, as described below.

\begin{figure*}[h!]
\centering
\includegraphics[width=0.48\textwidth,clip]{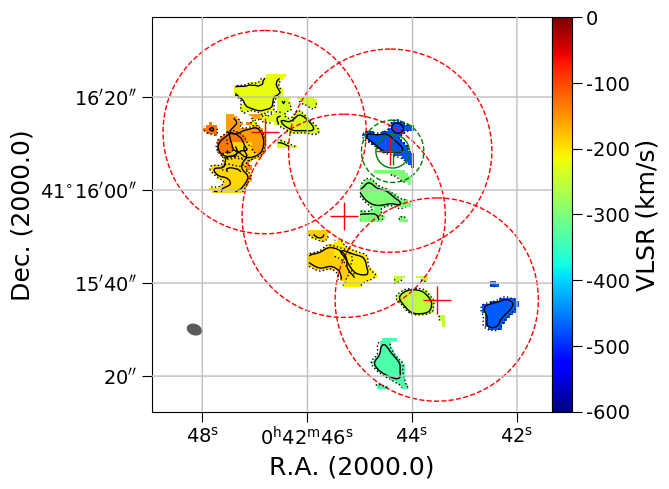}      
\includegraphics[width=0.48\textwidth,clip]{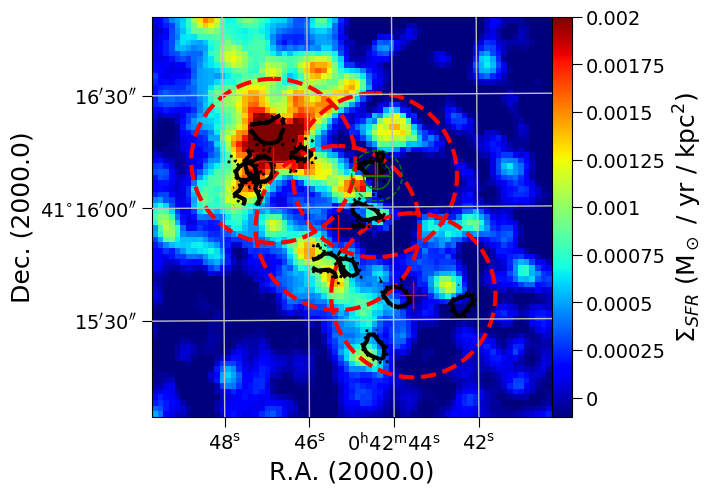}      
\caption{Left: Mean velocity map of the selected clumps with 1$\sigma$(dotted) and 2$\sigma$ (solid) intensity contours around clump. Right: Map of the surface density star formation rate in M31 central region from \citet{2013ApJ...769...55F} with the contours of our clumps superposed.}
\label{fig:CloudMapsVelo}
\end{figure*}

\subsection{Principal Component Analysis}
\label{ssect:PCA}
We apply PCA to the full sample of molecular cloud candidates (both negative and positive clumps), using the following parameters: $SNR_\text{peak}$, $SNR_\text{tot}$, $R_\text{eq}$ and $\sigma_\text{v}$.
These four parameters are first standardized as follow: $\widetilde{T} = (T- \langle T \rangle)/\sigma_T$, where T is the studied parameter, $\langle T \rangle$ its mean value and $\sigma_T$ its standard deviation.
Second, with the Python \textsc{ScikitLearn} library, we  extract two principal components, corresponding to the eigenvectors with the highest eigenvalues. They are a linear combination of these four standardized parameters:
\begin{equation}
\label{eq:PCA}
\begin{split}
PC1 &= 0.23 \, \widetilde{R_\text{eq}} + 0.57 \, \widetilde{SNR_\text{peak}} + 0.67 \widetilde{SNR_\text{tot}} + 0.41 \widetilde{\sigma_\text{v}}\\
PC2 &= 0.92 \, \widetilde{R_\text{eq}} - 0.38 \, \widetilde{SNR_\text{peak}} + 0.01 \widetilde{SNR_\text{tot}} + 0.001 \widetilde{\sigma_\text{v}}
\end{split}
\end{equation}
Figure \ref{fig:PCA} maps the sample on two axes defined as the two main principal components accounting for 50\,$\%$ and 25\,$\%$ of the  behaviour of the sample. The top panel maps the clumps kept by the cleaning procedure, while the bottom panel displays the identified side lobes and negative clumps. After visual inspection and cross-check with the negative signals, we eliminate the red dots in the top panel of Figure \ref{fig:PCA} below the blue dashed line. Hence, on the top panel, we keep blue dots with:
\begin{equation}
1.15 \, \widetilde{R_\text{eq}} + 0.68 \, \widetilde{SNR_\textrm{tot}} + 0.41 \, \widetilde{\sigma_\textrm{v}} + 0.19 \, \, \widetilde{SNR_\textrm{peak}} \ge 0.55
\end{equation}
The bottom panel displays candidates rejected as side lobes and/or exhibiting a negative signal. This shows the limitation of this analysis. Amplitude of side lobes can be important and 4 positive side lobes with relatively high significance with respect to the selected signals (in the top panel) are rejected.

We thus keep 12 clumps whose parameters are displayed in Table \ref{tab:AllDat}. They have SNR$_\mathrm{peak}\ge 4.2$ and SNR$_\mathrm{tot}\ge 15.2$ as displayed in Figure  \ref{fig:snrtot}. The maps and corresponding spectra are displayed in Figure \ref{fig:Maps}.


\subsection{Final sample selected}
\label{ssec:FinalSelec}
We show the maps and integrated spectrum for all 12 selected clumps in Figure~\ref{fig:Maps}, by order of decreasing $SNR_\text{tot}$. As said in Section~\ref{ssect:3Dsize}, the position angle of several clumps differs significantly from the position angle of the beam. It tends to happen for extended clumps which display structures.

\begin{table*}
\centering
\caption{Properties of the 12 selected molecular clumps extracted from our datacube. The offsets are calculated according to the optical centre of M31 at coordinates at J2000 coordinates $00h\,42^m\,44.37^s$ and $+41^{\circ}\,16\arcmin\,8.34\arcsec$ \citep{1992ApJ...390L...9C}. $V_0$ is the velocity at the peak flux $\Phi_{\text{max}}$ and the noise of the associated spectrum is $\sigma_{\text{noise}}$. The S/N ratio $SNR_{\text{peak}}$ is derived from these. $\sigma_{\text{v}}$, $R_{\text{eq}}$ and $F_{\text{CO}}$ are the extrapolations found through the methodology in Sect.~\ref{ssect:3Dsize}. $SNR_{\text{tot}}$ is found with $F_{\text{CO}}$ and the noise of the integrated spectrum associated.}
	\begin{tabular}{cccccc|ccccc}
	\hline \hline 
\multicolumn{6}{c}{} & \multicolumn{5}{c}{Extrapolations}  \\
ID & Offsets & $V_0$ & $\Phi_{\text{max}}$ & $\sigma_{\text{noise}}$ & $SNR_{\text{peak}}$ & $\sigma_\text{maj}$ & $\sigma_\text{min}$ &  $\sigma_{\text{v}}$ & $F_{\text{CO}}$ & $SNR_{\text{tot}}$\\
 & $\left(arcsec\right)$ & $\left(km\,s^{-1}\right)$ & $\left(\frac{mJy}{beam}\right)$ & $\left(\frac{mJy}{beam}\right)$ & & $\left(pc\right)$ & $\left(pc\right)$ & $\left(km\,s^{-1}\right)$ & $\left(Jy\,km\,s^{-1}\right)$ &  \\\hline\hline 
 1 & $-5.2,-32.5$&  45.1$\pm$0.7 & 55.8 & 5.4 & 10.3 &  6.6$\pm$0.4 &  5.7$\pm$0.3 &  8.8$\pm$0.3 & 2.23 $\pm$ 0.09 & 64.4 \\
 2 & $ 30.8,2.3$   & 145.5$\pm$0.4 & 58.4 & 4.5 & 12.9 &  6.0$\pm$0.4 &  5.8$\pm$0.4 &  6.0$\pm$0.3 & 1.48 $\pm$ 0.09 & 56.9 \\
 3 & $ 34.4,-3.8$  & 110.9$\pm$1.5 & 26.8 & 5.6 &  4.8 &  9.7$\pm$0.5 &  7.6$\pm$0.4 &  7.8$\pm$0.2 & 1.42 $\pm$ 0.05 & 49.4 \\
 4 & $ 29.0,10.8$  &  76.6$\pm$0.7 & 34.8 & 3.7 &  9.4 &  8.1$\pm$0.4 &  5.9$\pm$0.3 &  6.0$\pm$0.2 & 1.10 $\pm$ 0.06 & 46.4 \\
 5 & $ 13.1,-22.7$ & 112.9$\pm$1.2 & 29.9 & 5.1 &  5.9 &  8.7$\pm$0.5 &  5.5$\pm$0.3 &  6.7$\pm$0.2 & 1.15 $\pm$ 0.05 & 39.6 \\
 6 & $  0.9,-45.9$ & -35.8$\pm$0.9 & 48.1 & 7.7 &  6.3 &  5.4$\pm$0.3 &  5.4$\pm$0.3 &  6.5$\pm$0.2 & 1.98 $\pm$ 0.07 & 36.0 \\
 7 & $ 19.8,5.9$   &  65.8$\pm$1.4 & 20.5 & 4.1 &  5.0 &  9.2$\pm$0.7 &  4.3$\pm$0.3 & 10.7$\pm$0.3 & 0.73 $\pm$ 0.03 & 31.1 \\
 8 & $ 35.7,0.4$   & 171.5$\pm$1.5 & 27.7 & 5.0 &  5.5 &  7.3$\pm$0.4 &  7.3$\pm$0.4 &  8.7$\pm$0.3 & 0.75 $\pm$ 0.04 & 23.8 \\
 9 & $  2.1,-9.9$  &   4.3$\pm$1.5 & 15.5 & 3.2 &  4.8 & 11.4$\pm$0.5 &  7.9$\pm$0.3 &  4.4$\pm$0.2 & 0.44 $\pm$ 0.02 & 22.8 \\
10 & $  7.6,-24.0$ &  95.8$\pm$1.0 & 21.4 & 5.1 &  4.2 & 10.7$\pm$0.6 & 10.6$\pm$0.5 &  6.2$\pm$0.2 & 0.65 $\pm$ 0.03 & 19.0 \\
11 & $-22.3,-33.7$ &-167.6$\pm$1.4 & 24.0 & 5.7 &  4.2 &  9.9$\pm$0.4 &  9.5$\pm$0.3 &  3.4$\pm$0.2 & 0.57 $\pm$ 0.03 & 15.4 \\
12 & $  1.5,  1.7$ &-176.5$\pm$1.1 & 19.1 & 3.5 &  5.4 & 10.8$\pm$0.7 &  7.8$\pm$0.5 &  6.1$\pm$0.2 & 0.39 $\pm$ 0.02 & 15.2 \\
    \hline \hline  
	\end{tabular}
    \label{tab:AllDat}
\end{table*}

\begin{figure}[h!]
\centering
\includegraphics[width=0.4\textwidth,clip]{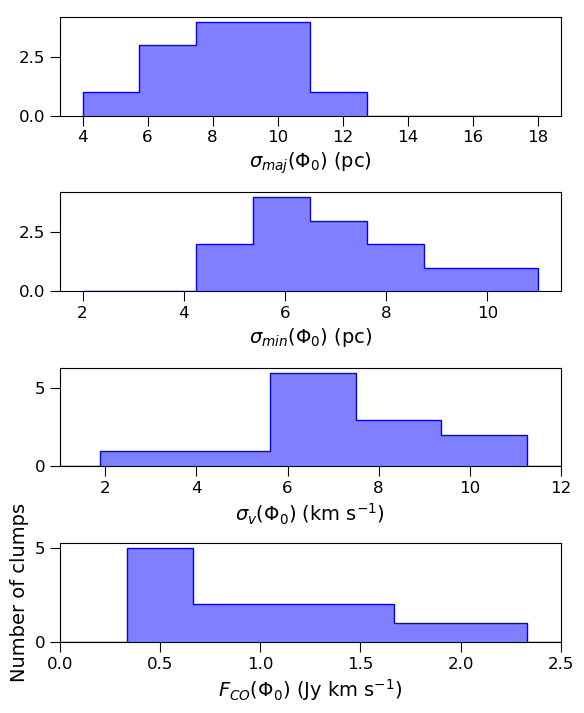}      
\caption{Distribution of the $\sigma_\text{maj}$, $\sigma_\text{min}$, $\sigma_\text{v}$ and F$_\text{CO}$ for the selected clumps. We find the average values : $\langle\sigma_\text{maj}\rangle$ = 8.6 $\pm$ 2.5 pc, $\langle\sigma_\text{min}\rangle$ = 7.0 $\pm$ 1.7 pc, $\langle\sigma_\text{v}\rangle$ = 6.8 $\pm$ 1.1 km s$^{-1}$ and $\langle F_\text{CO}\rangle$ = 1.1 $\pm$ 0.6 Jy km s$^{-1}$.}
\label{fig:DistribSigmas}
\end{figure}

Our selection went through a 5-step procedure:

\begin{enumerate}
\item In Sect.~\ref{ssect:core}, the 3-$\sigma$ peak detection allowed us to find 54 positive candidates within the data-cube, including faint signals which needed further investigation. At this stage, 47 negative clumps were kept in order to perform a statistical analysis:  those negative (false) detection with similar characteristics compared to our sample of positive candidates provide statistical information on the noise affecting the data. The SNR$_\textrm{peak}$ of each core was calculated.
\item In Sect.~\ref{ssect:3Dsize}, we extrapolated the values of $\sigma_\textrm{maj}$, $\sigma_\textrm{min}$, $\sigma_\textrm{v}$ and F$_\textrm{CO}$ for both positive and negative clumps. This allowed us to find the values for R$_\textrm{eq}$ and SNR$_\textrm{tot}$.
\item In Sect.~\ref{ssec:velo}, with the use of \textsc{Gildas Clean} algorithm, we check clumps with identical velocities for side lobes. Positive and negative clumps, some with a significant SNR$_\textrm{tot}$ were found to be side lobes of three positive 3-$\sigma$ clumps.
\item In Sect \ref{ssect:PCA}, we apply a statistical analysis on our sample of clumps. 
\end{enumerate}

Hence, we complete the selection procedure. The final map is presented in Figure~\ref{fig:CloudMapsVelo}, with the clump mean velocity in colour. Although the CO emission is patchy, it follows the general velocity field found in H$\alpha$ \citep{Boulesteix1987}.
The CO emission globally corresponds well to the star formation map of \citet{2013ApJ...769...55F}.
12 clumps are kept with $SNR_\text{tot}$ showed in Figure~\ref{fig:snrtot}. Their properties are summed up in Table~\ref{tab:AllDat}. The distribution of $\sigma_\text{maj}$, $\sigma_\text{min}$, $\sigma_\text{v}$ and F$_\text{CO}$ for selected clumps is displayed in Figure~\ref{fig:DistribSigmas}.

\begin{figure}[h!]
    \centering
    \includegraphics[width=0.48\textwidth,clip]{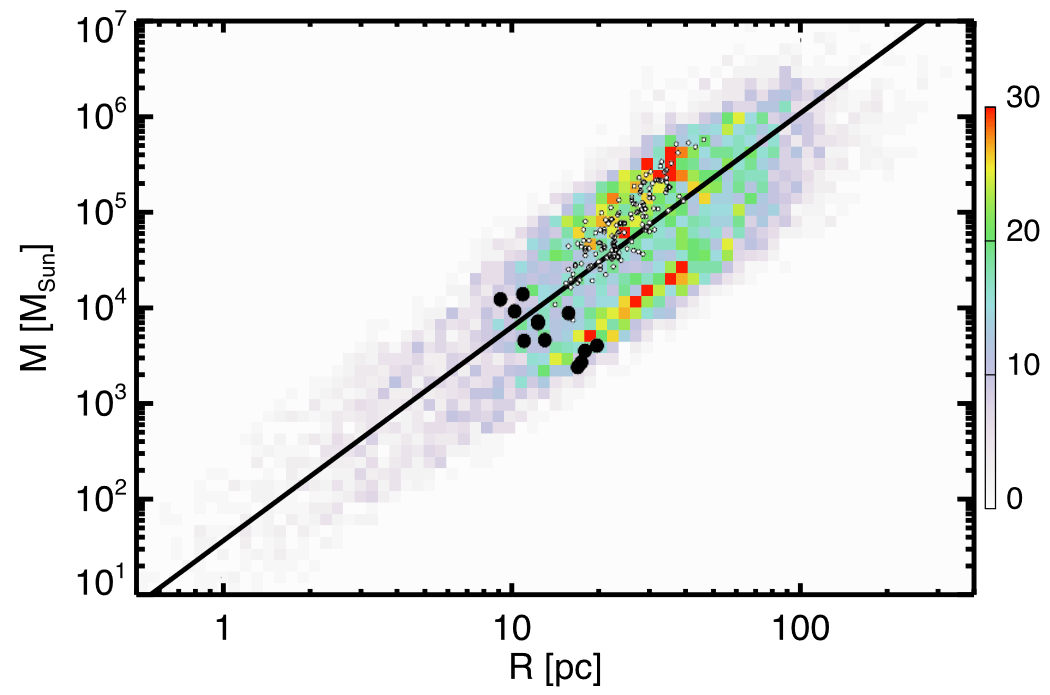}
    \caption{Mass-size relation for the M31 circum-nuclear clumps (large full black symbols) superposed on the relation obtained for Milky Way clouds by \cite{Miville2017}. The colour scale is proportional to the density of points. The solid line is a linear fit of slope 2.2. The small empty symbols are the clouds identified in Centaurus A northern filament by \cite{Salome2017AA}.}
    \label{fig:larson}
\end{figure}

\begin{figure}[h!]
    \centering
    \includegraphics[width=0.48\textwidth,clip]{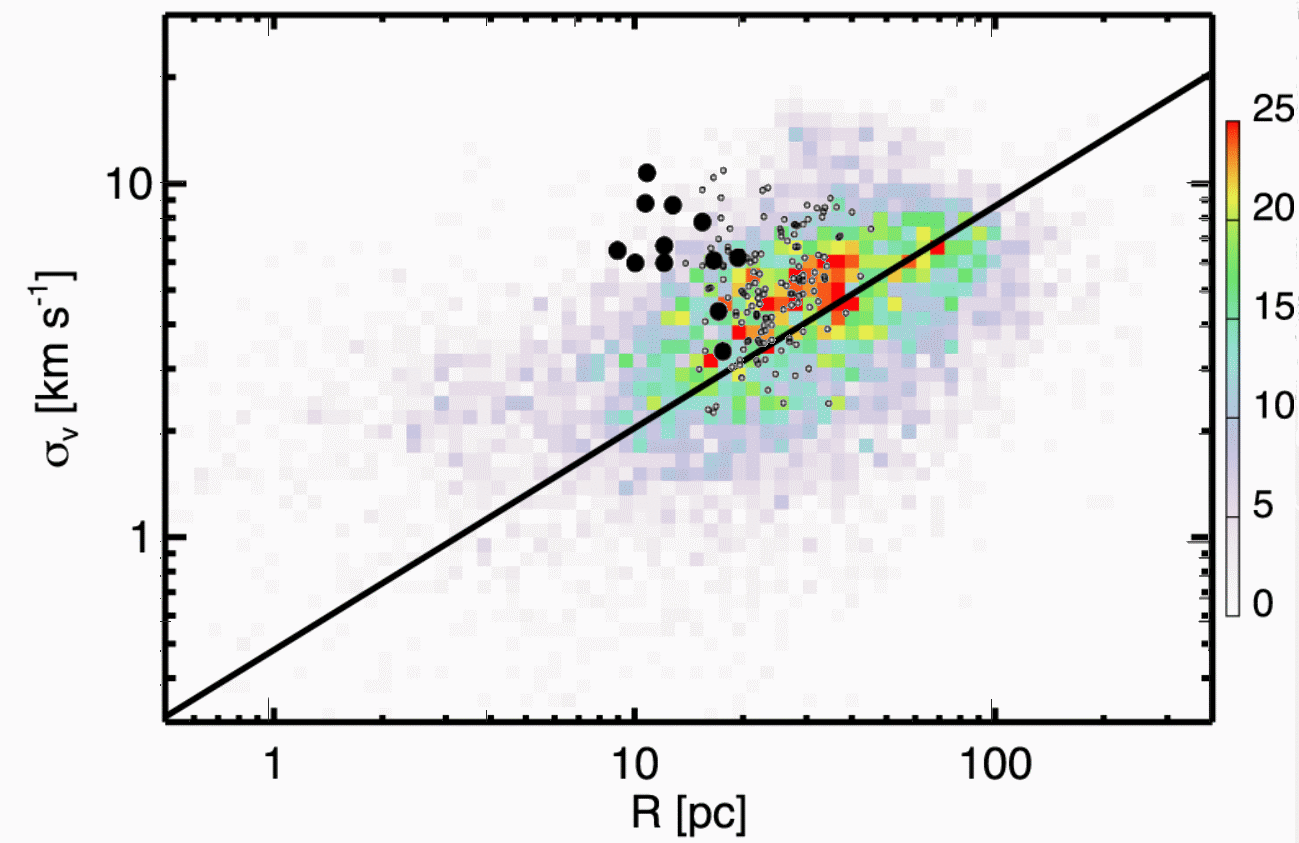}
    \caption{Velocity dispersion-size relation for the M31 circum-nuclear clumps (large full black symbols) superposed on the relation obtained for Milky Way clouds by \cite{Miville2017}. Like in Figure \ref{fig:larson}, the colour scale is proportional to the density of points. The solid line is a linear fit of slope 0.63. The small empty symbols are the clouds identified in Centaurus A northern filament by \cite{Salome2017AA}. We observe a clear observational bias: we only detect the clumps with the largest velocity dispersion.}
    \label{fig:larson2}
\end{figure}
\begin{table*}
 \centering
\caption{Characteristics of the 12 selected molecular clumps. The descriptive properties are shown in Table~\ref{tab:AllDat}. The offsets are calculated according to the optical centre of M31 at J2000 coordinates $00^h\,42^m\,44.37^s$ and $+41^{\circ}\,16\arcmin\,8.34\arcsec$. The calculation of R$_\textrm{eq}$ is explained in Sect.~\ref{ssect:3Dsize}. The CO luminosity $L_{\text{CO}}'$, the luminous mass $M_{\text{H}_2}$, the surface density $\Sigma_{\text{H}_2}$, the virial parameter $\alpha_{\text{vir}}$ and the H$_2$ column density N$_{\text{H}_2}$ calculations are detailed in Sect.~\ref{sect:clprop}.}
\begin{tabular}{cccccccc}
	\hline \hline 
ID & Offsets & R$_\textrm{eq}$ & $L_{\text{CO}}'$ & $M_{\text{H}_2}$ & $\Sigma_{\text{H}_2}$ & $\alpha_{\text{vir}}$ & N$_{\text{H}_2}$ \\
   & $\left(arcsec\right)$ & pc & $K\,km\,s^{-1}\,pc^2$ & $10^3 M_{\odot}$ & $M_{\odot}\text{pc}^{-2}$ & & $10^{20}\,cm^{-2}$ \\\hline\hline 
 1 & $-5.2,-32.5$  & 11.0 $\pm$ 0.7 & 3321 $\pm$ 138 & 14.5 $\pm$ 0.6 & 36.4 $\pm$ 6.3 &  68 $\pm$ 11 & 20.2 $\pm$ 3.5 \\
 2 & $30.8,2.3$    & 10.3 $\pm$ 0.9 & 2202 $\pm$ 137 &  9.6 $\pm$ 0.6 & 27.8 $\pm$ 6.5 &  44 $\pm$ 10 & 15.4 $\pm$ 3.6 \\
 3 & $34.4,-3.8$   & 15.8 $\pm$ 0.7 & 2119 $\pm$  77 &  9.2 $\pm$ 0.3 & 11.3 $\pm$ 1.4 & 121 $\pm$ 15 &  6.2 $\pm$ 0.8 \\
 4 & $28.9,10.8$   & 12.4 $\pm$ 0.6 & 1644 $\pm$  84 &  7.2 $\pm$ 0.4 & 14.3 $\pm$ 2.1 &  71 $\pm$ 13 &  7.9 $\pm$ 1.2 \\
 5 & $13.1,-22.7$  & 12.4 $\pm$ 0.5 & 1709 $\pm$  70 &  7.5 $\pm$ 0.3 & 14.9 $\pm$ 1.9 &  87 $\pm$ 12 &  8.3 $\pm$ 1.1 \\
 6 & $0.9,-45.9$   &  9.2 $\pm$ 0.7 & 2940 $\pm$ 107 & 12.8 $\pm$ 0.5 & 46.6 $\pm$ 8.4 &  35 $\pm$  6 & 25.9 $\pm$ 4.7 \\
 7 & $19.8,5.9$    & 11.1 $\pm$ 0.5 & 1087 $\pm$  49 &  4.7 $\pm$ 0.2 & 11.8 $\pm$ 1.7 & 311 $\pm$ 45 &  6.6 $\pm$ 0.9 \\
 8 & $35.7,0.4$    & 13.1 $\pm$ 0.8 & 1112 $\pm$  65 &  4.8 $\pm$ 0.3 &  8.6 $\pm$ 1.6 & 236 $\pm$ 47 &  4.7 $\pm$ 0.9 \\
 9 & $2.1,-9.9$    & 17.5 $\pm$ 0.6 &  650 $\pm$  30 &  2.8 $\pm$ 0.1 &  2.8 $\pm$ 0.3 & 142 $\pm$ 23 &  1.5 $\pm$ 0.2 \\ 
10 & $7.6,-24.0$   & 19.8 $\pm$ 1.1 &  968 $\pm$  41 &  4.2 $\pm$ 0.2 &  3.3 $\pm$ 0.5 & 209 $\pm$ 30 &  1.8 $\pm$ 0.3 \\
11 & $-22.3,-33.7$ & 18.0 $\pm$ 0.7 &  852 $\pm$  40 &  3.7 $\pm$ 0.2 &  3.5 $\pm$ 0.4 &  66 $\pm$ 12 &  1.9 $\pm$ 0.2 \\
12 & $1.5,1.7$     & 17.0 $\pm$ 0.9 &  579 $\pm$  37 &  2.5 $\pm$ 0.2 &  2.7 $\pm$ 0.4 & 294 $\pm$ 56 &  1.5 $\pm$ 0.2 \\
\hline \hline 
\end{tabular}
\label{tab:AllDat2}
\end{table*}

\section{Cloud properties}
\label{sect:clprop}
 In this Section, our goal is to determine the physical properties of the identified molecular clouds in the centre of M31, size, velocity dispersion, mass and Virial ratio, to compare
them to nearby galaxy clouds, and to their scaling relations.
 
\subsection{Velocity dispersion}
\label{ssect:VelDisp}
The average velocity dispersion of our selected sample, computed
as in Equation (4) is $6.8 \pm 1.3 \, km\,s^{-1}$. The corresponding histogram is shown
in Figure \ref{fig:DistribSigmas}.  
\citet{Calduprimo2016} detected molecular clouds, mostly located in the 10 kpc ring of M31, found a median velocity dispersion of $3.1 \pm 1.1 \, km \, s^{-1}$ with interferometric data.

We do find here a median velocity dispersion of $ 6.35 \, km \, s^{-1}$. However, the relation velocity dispersion and radius do not follow the expected Larson's relation (cf Figure \ref{fig:larson2}), contrary to the mass-radius relation, as discussed below.
While averaging over the whole extent of the clouds, it is possible to find 
a dispersion as low as 3 km/s in one clump (note that the corresponding FWHM would be $\sim$ 7 km/s), the present spectral resolution (and sampling) (FWHM of 5.07 km/s) is a strong limitation in our determinations. 

\subsection{Mass estimates}
\label{ssect:mass}
The luminous mass $M_{\text{H}_2}$ of the clumps can be derived from the total integrated flux. First we calculate the CO luminosity L$_\text{CO}'$, which is defined as \citep{Solomon1997}:
  \begin{equation}
  \frac{L_\text{CO}'}{K\,km\,s^{-1}\,pc^2} = \frac{3.25 \times 10^7}{\left(1+z\right)^3} \, \frac{F_\text{CO}}{Jy.km.s^{-1}} \, \left(\frac{D_\text{L}}{Mpc}\right)^2 \, \left(\frac{\nu_\text{rest}}{GHz}\right)^{-2},
  \label{eq:lcop}
  \end{equation}
where $D_\text{L} = 780 \, kpc$ is the distance to M31, $\nu_\text{rest} = 115.271$\,GHz is the observed frequency and $z = 0$. This leads to the luminous mass:
  \begin{equation}
  \frac{M_{\text{H}_2}}{M_{\odot}} = 4.4\frac{L_\text{CO}'}{K\,km\,s^{-1}\,pc^2}\frac{X_\text{CO}}{2\times10^{20}\,cm^{-2}\left(K\,km\,s^{-1}\right)^{-1}},
  \label{eq:mlum}
  \end{equation}
where X$_\textrm{CO}$ is the $CO$-to-$H_2$ conversion
factor \citep{2013ARA&A..51..207B}. The mass is then given by $M_{\textrm{H}_2} = \alpha_\textrm{CO}\, L_\textrm{CO}'$, with $\alpha_\textrm{CO} = 4.36\,M_{\odot}\,\left(K\,km\,s^{-1}\,pc^{2}\right)^{-1}$, based on the 
conversion factor of the Milky Way $3.2\,M_{\odot}\,\left(K\,km\,s^{-1}\,pc^2\right)^{-1}$  corrected by a 
factor 1.36 to account for interstellar helium \citep{Tacconi2010}. We find a total H$_2$ mass of $\left(8.4 
\pm 0.4\right) \times 10^4 \, M_\odot$ which is coherent with \citet{2013A&A...549A..27M} estimation of a 
minimum total mass of $4.2 \times 10^4 \, M_\odot$. We have a mean $M_{\text{H}_2}$ of $\left(7.0 \pm 
0.3\right) \times 10^3 \, M_\odot$ for the detected clouds.
We find our values for the mass and size of clumps to be consistent with Larson's mass -- size relation found in \citet{Salome2017AA}, \citet{Gratier2012} and \citet{Rosolowsky2007}, cf Figure \ref{fig:larson}. It is not the case for the velocity -- size and velocity -- mass relations though.
 This might be explained in part by the fact that our minimal value of $\sigma_\text{v}$ is biased high by our spectral resolution and selection process (we ask for a minimum of two consecutive channels). Hence, our velocity dispersions are
somewhat over-estimated. Although, for signals  with SNR$_{peak} <10$, \citet{Rosolowsky2006} has shown that measurements based on interferometric data tend to underestimate size and velocity dispersion.

The surface density $\Sigma_{\text{H}_2}$ is found by dividing the molecular mass of the clump by its area in squared parsecs. The area is calculated for an elliptic shape with the major and minor axis deconvolved from
the beam. We find an average surface density of $15.3 \pm 2.6 \, M_\odot\,pc^2$. This is three times less than in NGC 5128, about two times less than in the Milky Way but close to the value found in the LMC, $21 \pm 9 \, M_\odot\,pc^2$ \citep{Salome2017AA,Miville2017,Hughes2013}.

The molecular column density is calculated as follows :
  \begin{equation}
  N_{\text{H}_2} = \frac{\Sigma_{\text{H}_2}}{M_\odot pc^2} \frac{M_\odot}{kg} \left(\frac{2.3 m_p}{kg}\right)^{-1} \left(\frac{pc}{cm}\right)^{-2},
  \label{eq:nh2}
  \end{equation}
where we multiply the proton mass $m_p$ by $2.3$ to take into account the helium mass. The average column density is $\left(8.5 \pm 1.5\right) \times 10^{20} \, cm^{-2}$
  
In order to estimate the impact of the magnetic, kinetic and gravitational energy of the clumps, we finally calculate the Virial parameter \citep{Bertoldi1992ApJ,McKee1992AJ}:
  \begin{equation}
  \alpha_\text{vir} = 5 \sigma_\text{v}^2 \frac{R_\text{eq}}{G M_{\text{H}_2}}.
  \label{eq:virpar}
  \end{equation}
A value of $\alpha_\text{vir} \leq 1$ indicates a gravitationally bound core with a possible support of magnetic fields while if $\alpha_\text{vir} \leq 2$, it means the core is gravitationally bound without magnetic support. 
In our case, clumps have $\alpha_\text{vir} \gg 2$, their kinetic energy appears dominant and they are not virialized. 
Compared to the virial parameter in \citet{Salome2017AA,Miville2017} and \citet{Hughes2013} for giant molecular clouds, our mean value is $140 \pm 72$ which is one to two orders of magnitude higher than these 
since our clumps are smaller and less massive.

Figure \ref{fig:larson2} reveals that the velocity dispersions of our
 measured clumps lie well above the relation for Milky Way clouds. Certainly our spectral resolution prevents us to detect velocity dispersions below the relation.
 However, even after deconvolution from the spectral resolution, the majority of clumps have a broad dispersion. We conclude that the virial factors $\alpha_{vir}$ are in a large majority much larger than 1, and the clumps are not virialized. Certainly, they are transient agglomerations of smaller entities, which might be virialized.

\begin{figure*}[h!]
\centering
\includegraphics[width=0.33\textwidth,clip]{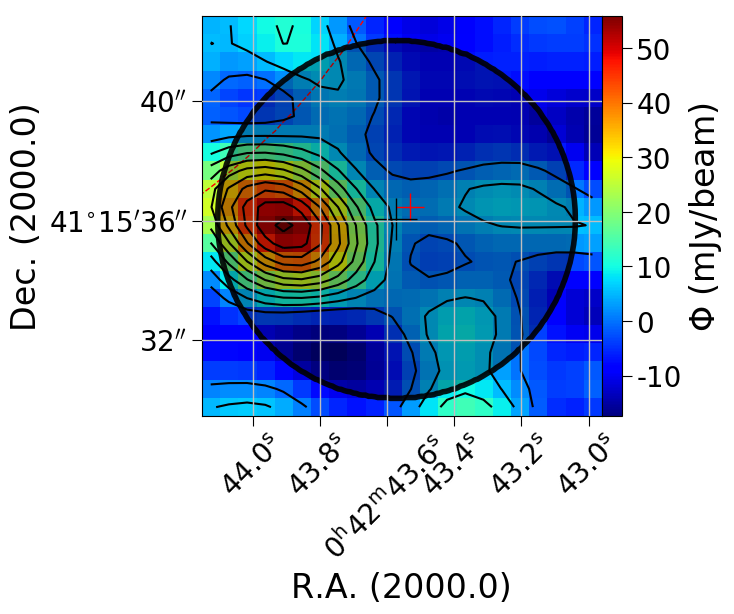}      
\includegraphics[width=0.33\textwidth,clip]{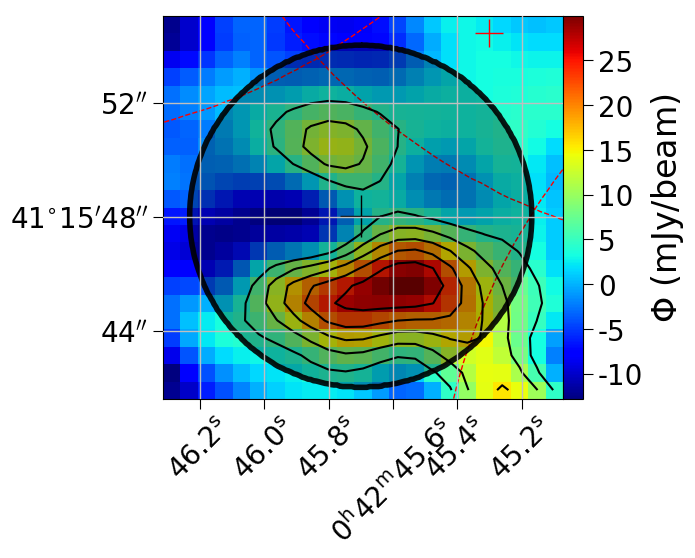}      
\includegraphics[width=0.33\textwidth,clip]{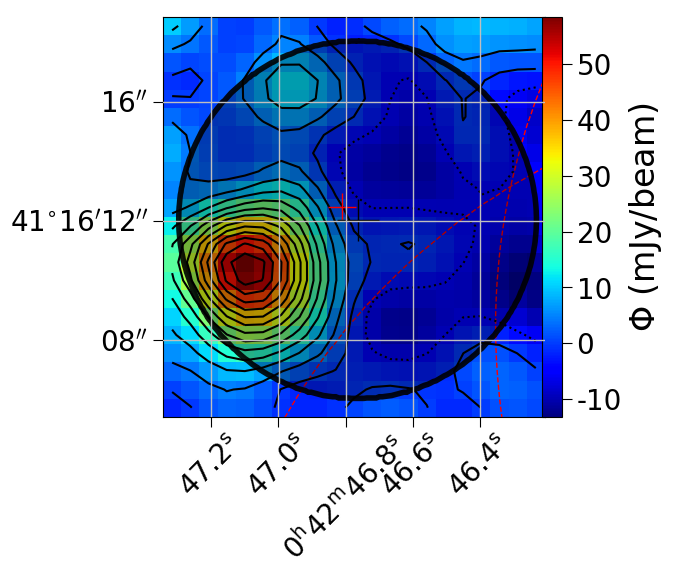}      
\caption{The PdB channels maps associated to the positions displayed in  Table~\ref{tab:depletion}. The darkened region represents beam of the single dish observation from \citet{2013A&A...549A..27M}. We notice a position offset and that the whole clump is not always contained in the single dish observations, which explains the missing flux. Velocities obtained with HERA IRAM-30m observations do correspond to the velocities measured with PdB IRAM interferometric data.}
\label{fig:vignettes}
\end{figure*}

\begin{figure}[h!]
\centering    
\includegraphics[width=0.48\textwidth,clip]{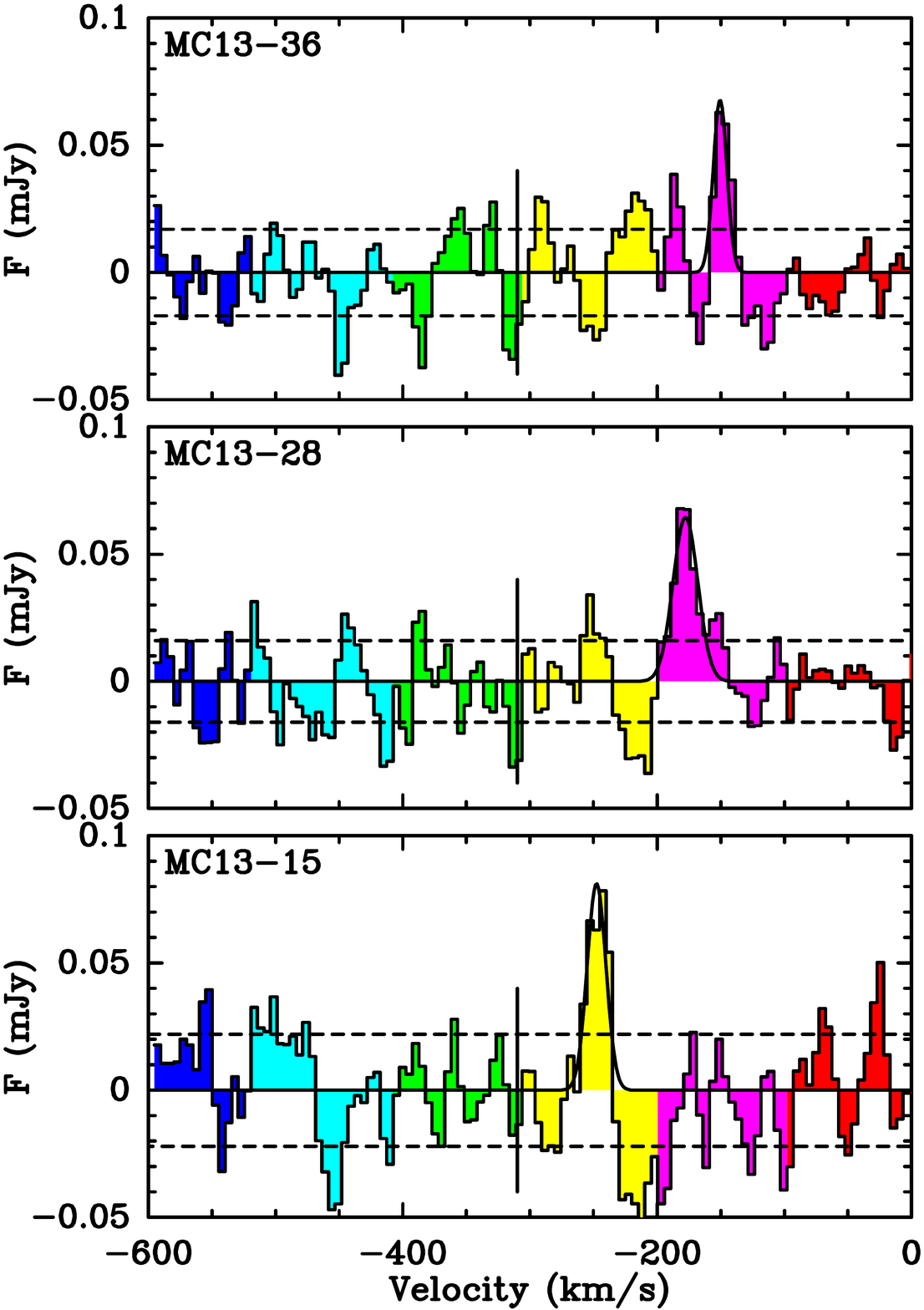}      
\caption{CO(1-0) spectra corresponding to the CO(2-1) detections from \citet[][hereafter MC13]{2013A&A...549A..27M}. The PdB IRAM interferometric data has been reprojected on the grid of the data cube obtained with HERA on IRAM 30-m telescope as described in MC13. We show here the spectra in the 12-arcsec beams displayed in the Figure \ref{fig:vignettes}. A vertical line indicates the systemic velocity. In each panel, the best Gaussian fit used to integrate the CO line, and which results are displayed in Table \ref{tab:depletion}.}
\label{fig:3spectra}
\end{figure}

\section{Discussion}
\label{sect:disc}
In the following, we discuss how this new molecular gas detection compares with other measurements. In Section \ref{ssect:co}, we discuss the line ratios of some clumps based on CO(2-1) detection from \citet{2013A&A...549A..27M} and CO(3-2) from \citet{2018arXiv181210887L}. In Sect. \ref{ssect:TSF}, we study the spectral energy distribution of this region and discuss the star formation tracers available in this region. We show the absence of correlation between dust, most probably heated by the interstellar radiation field, and the FUV due to the old bulge stellar population plus the 200\,Myr inner disc discussed by \citet{2012ApJ...745..121L}.  Last, relying on the SFR estimate of this region,  we discuss, in Sect. \ref{ssect:KSLaw}, the extreme position of our measurement on the Kennicutt-Schmidt law.
\subsection{CO(2-1)-to-CO(1-0) line ratio}
\label{ssect:co}
\subsubsection{Data}
\label{sssect:co}

\begin{table*} 
    \begin{tabular}{l|cc|cc|cc|cc|cc|cc}
	\hline \hline 
MC13 : Offsets & \multicolumn{2}{c}{$S_{CO}\Delta v$}  & \multicolumn{2}{c}{$V_0$}& \multicolumn{2}{c}{$\Delta V$} & \multicolumn{2}{c|}{S$_{peak}$} & \multicolumn{2}{c}{rms }& \multicolumn{2}{c}{(2-1)/(1-0)}\\  
ID : arcsec & \multicolumn{2}{c|}{Jy\,km\,s$^{-1}$}  & \multicolumn{2}{c|}{km\,s$^{-1}$}& \multicolumn{2}{c|}{km\,s$^{-1}$} & \multicolumn{2}{c}{mJy} & \multicolumn{2}{c}{mJy} & Flux & Temp.\\ 
&  (2-1) & (1-0)& (2-1) & (1-0)& (2-1) & (1-0)& (2-1) & (1-0)& (2-1) & (1-0) & ratio & ratio\\ \hline
15 : \,-9.0,-32.3 & 3.0$\pm$0.4 & 1.5$\pm$0.3& -248$\pm$2 & -248. & 24$\pm$4 & 17$\pm$3 & 115 & 81 & 16 & 22 & 1.41 & 0.35$ \pm 0.08$\\
28 : +15.,-20.3 & 1.8$\pm$0.4 & 1.6$\pm0.3$ & -190$\pm$1 & -178 & 14$\pm$3 & 23$\pm$6& 130 & 64 & 19 & 16& 2.03& 0.51$ \pm 0.15$ \\
36 : +27.,+3.7 & 4.7$\pm$0.8 & 0.91$\pm$0.2& -153$\pm$5 & -150.& 53$\pm$11   & 13$ \pm$3 & 84 & 68. & 20.5 & 17& 1.23 & 0.31$\pm 0.09$\\ \hline
\end{tabular}
\caption{Characteristics of the CO(2-1) and CO(1-0) lines integrated in a 12$^{\prime\prime}$ FWHM beam at 3 positions. The MC13 corresponds to the identification proposed in \citet{2013A&A...549A..27M}, the offsets of each position are computed with respect to the optical centre. As the parameters, namely integrated flux $F_{CO} = S_{CO}\Delta v$, the central velocity $V_0$, the linewidth $\Delta V$, the peak value S$_{peak}$, the rms level of noise computed on the baseline, and the line ratios computed in Jy and in K, are derived from a Gaussian fit in CLASS in the GILDAS environment.}
\label{tab:depletion}
\end{table*}

\begin{table*} 
    \begin{tabular}{c|cc|ccccccc|cccc|c}
	\hline \hline 
	MC13 & $\eta^{5K}_{bf}$ & $\eta^{30K}_{bf}$ & $T_C$  &  $T_W$ & $M_{dust}$   & $M_{star}$  & $\tau^{dust}_V$ & $\tau^{Diff}_V$ & SFR  & $U_{mean}$ & $U_{min}$ & $T_{mean}$ & $T_{min}$ & SFR\\\hline
  ID  &$10^{-2}$ & $10^{-2}$ & K &K  & $M_\odot$ & $10^7 M_\odot$ & & &$ M_\odot$/Myr & & & K & K & $ M_\odot$/Myr \\
    & & & 
      \multicolumn{7}{c|}{(Viaene et al. 2014)
      }
    & \multicolumn{4}{c|}{\citep{2014ApJ...780..172D}}& \multicolumn{1}{c}{(F13)}\\ \hline
	15 & 4.8 & 4.0 & 29.8 & 63.7 & 33 & 1.8 & 0.152 & 0.022 & 0.22 & 18.6 & 15.0 & 29.3& 28.3& 8\\
	28 & 3.1 & 2.7 & 29.8 & 63.7 & 33 & 1.8 & 0.152 & 0.022 & 0.22 & 32.3& 22.0 & 32.1& 30.13& 77 \\
	36 & 7.8 & 6.6  & 29.8 & 63.7 & 38 & 2.0 & 0.162 & 0.027 & 3.25 & 27.1 & 25.& 31.2 & 30.7 & 133\\ \hline
	\end{tabular}
	\caption{Characteristics derived by \citet{2014A&A...567A..71V} and \citet{2014ApJ...780..172D} from infrared data at the position of the CO(2-1) and CO(1-0) detections in $36^{\prime\prime} \times 36^{\prime\prime}$ and $18^{\prime\prime} \times 18^{\prime\prime}$ pixels. As described in Sect. \ref{ssect:co} and in \citet{2016A&A...585A..44M}, $\eta_{bf}$ provides the beam filling factor computed for a brightness temperature of 5\,K relying on the RADEX results presented in Sect \ref{ssect:co} and 30\,K assuming it is thermalised with the dust. $T_C$ and $T_W$ correspond to the cold and warm dust components, $M_{dust}$ and $M_{star}$ are the dust and stellar masses scaled to the prorata of the $12^{\prime\prime}$ CO FWHM beam. $\tau^{dust}_V$ and $\tau^{Diff}_V$ are the dust and diffuse dust optical depths, given for the order of magnitude as they are the poorest constrained parameters according to \citet{2014A&A...567A..71V}. SFR is the star formation rate scaled to the prorata of the  $12^{\prime\prime}$  CO  FWHM beam. The last column provides the SFR estimate of \citet{2013ApJ...769...55F} integrated on the 12$^{\prime\prime}$ beam.}
\label{tab:vignettes2}
\end{table*}

In \citet{2013A&A...549A..27M}, we analysed IRAM-30m CO(2-1) data obtained with the HERA instrument. These data overlap the IRAM-PdB CO(1$-$0) mosaic studied  here. We do not find a priori a one-to-one correspondence with the signals detected in 
\citet{2013A&A...549A..27M}. The CO(2-1) signal is detected in more regions than CO(1-0). However, many detections were below 4$\sigma$ and were not recovered by the IRAM-PdB data presented here. While this is not surprising given the low S/N ratio, the signal could be affected by interferometric filtering if the gas is extended. Indeed, \citet{Rosolowsky2006} studied that for a given mass, a significant fraction of the flux can ben lost when the  peak S/N is lower than 10. 3 spectra (labelled 15, 28 and 36 in \citet{2013A&A...549A..27M}), actually detected above 4$\sigma$ in CO(2-1), have a counter-part in the CO(1-0) interferometric data cube. Each CO(2$-1$) pointing corresponds to a 12-arcsec HPBW. 
These CO(1-0) maps of these 3 simultaneous CO(1-0) and CO(2-1) detections are displayed in Figure \ref{fig:vignettes} and correspond to the clumps N$^o$1, 2 and 5 in Table \ref{tab:AllDat}. As displayed in this figure, the CO(2-1) detections are not centred on the CO(1-0) peak of emission, and the CO(1-0) clumps do not fill the single dish beam.  In order to study the line ratio, we integrate interferometric data within 12" diameter circles at these 3 positions. We obtained the spectra displayed in Figure \ref{fig:3spectra}. We use the usual conversion factors 5\,Jy/K for the CO(2-1) single dish data and 24\,Jy/K for the CO(1-0) interferometric data.
The results of this analysis is presented in Table \ref{tab:depletion}.

In Table \ref{tab:vignettes2}, we provide the dust characteristics of the positions, where CO(1-0) and CO(2-1) are detected, derived from the maps of \citet{2014A&A...567A..71V}, \citet{2014ApJ...780..172D} and \citet{2013ApJ...769...55F}. There is a strong interstellar radiation field $U\sim20-30$ due to the bulge stellar population, while the  cold dust temperature is around 30\,K.

Following Eq. \ref{eq:lcop} and \ref{eq:mlum}, we estimate a mean luminous mass along the line of sight, and within the 12\,arcsec $=$ 40\,pc beams, for these 3 offsets of 8700\,M$_\odot$. The fraction of gas with respect to the stellar mass along these lines of sight is very small:  $4. \times 10^{-4}$, while the dust-to-gas ratio is about 0.01. This is a sign of high SFR recycling and high metallicity. The Andromeda circum-nuclear region lies well below the dust scaling relation presented by \citet{2018ARA&A..56..673G}, as this is observed for early-type galaxies.

\subsubsection{Evidences of local non-thermal equilibrium conditions}
\label{sssect:lte}
\begin{figure*}[h!]
\centering
\includegraphics[width=\textwidth,clip]{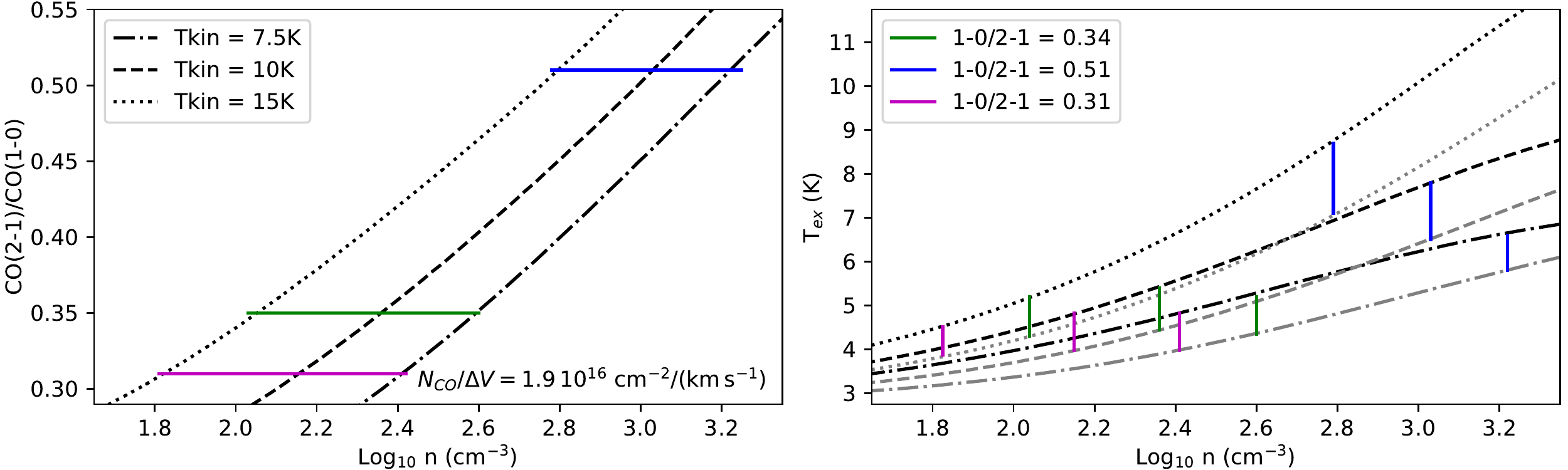}      
\caption{RADEX simulations. We consider a standard CO abundance N$_{CO} = 10^{-4} \langle N_{H_2} \rangle$. We measure a mean molecular hydrogen column density of $3 \times 10^{20}$\,cm$^{-2}$ that we correct for the surface beam filling factor ($\eta_{bf} = 0.052$) assuming the CO(2-1) is optically thick and computed for T$_{ex} =5$\,K. Last, we estimate a mean $\Delta V$ of 30\,km\,s$^{-1}$. The left panel displays the simulations for 3 different kinetic temperature $T_{kin}$. The left (resp. right) panels present the expected CO(2-1) to CO(1-0) line ratio (in temperature) (resp. the excitation temperature of the 2 lines) as a function of the hydrogen density. In the right panel, the black (resp. grey) lines correspond to the CO(1-0) (resp. CO(2-1)) line.  The green, blue and magenta ticks correspond to the line ratios measured in the 3 positions.  This shows that this gas is sub-thermally excited and well below the expected critical density for a kinetic temperature of 10\,K, $n^{1-0}_{crit} = 2.2\times 10^3$\,cm$^{-3}$ and $n^{2-1}_{crit} = 9.7\times 10^3$\,cm$^{-3}$.}
\label{fig:radex} 
\end{figure*}
\begin{figure}[h!]
\centering 
\includegraphics[width=0.48\textwidth,clip]{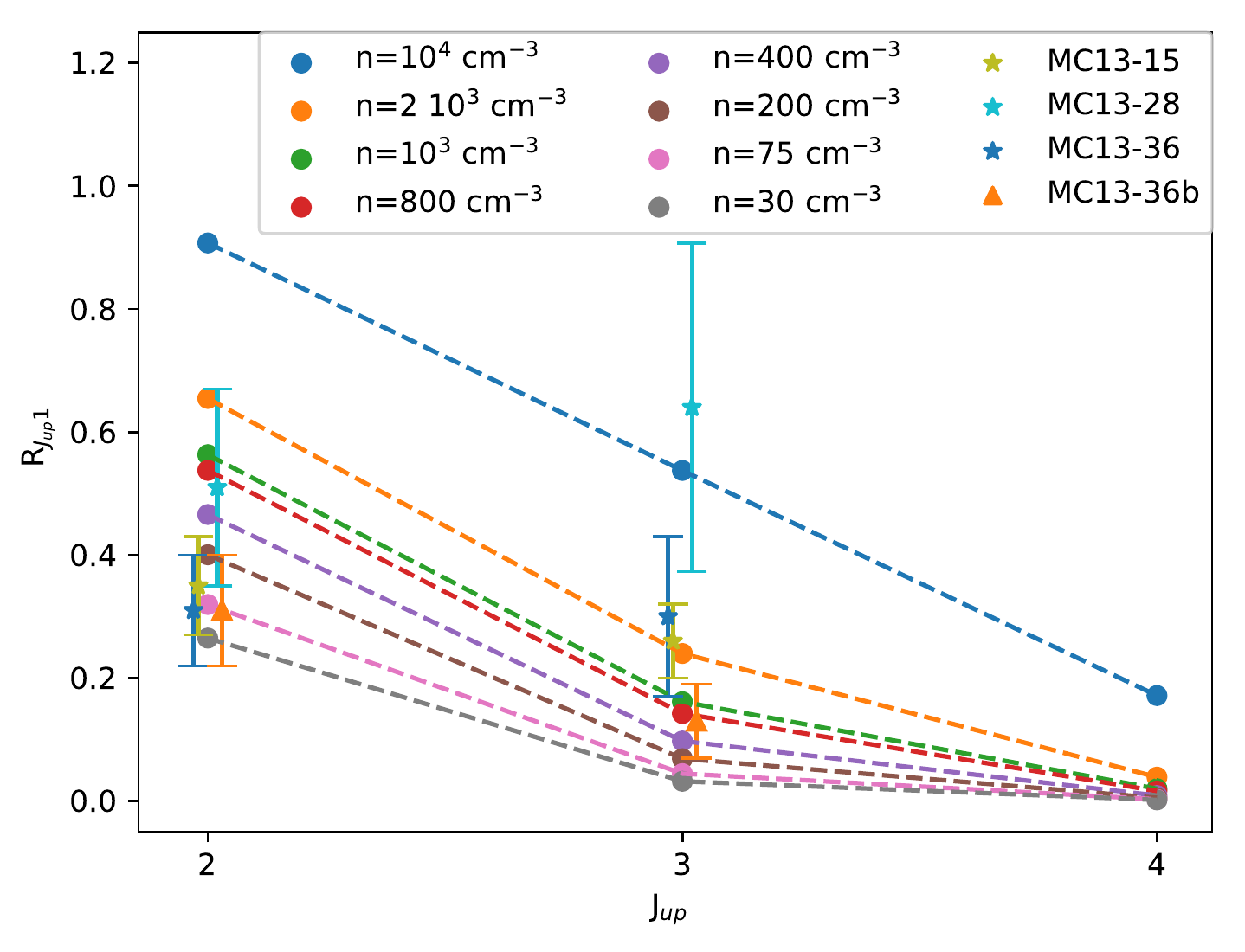}      
\caption{Comparison of the CO line ratios observed for 3 positions of the circumnuclear disc with RADEX simulations. As discussed in the text, we run RADEX simulations assuming a large velocity gradient, with a kinetic temperature of 15\,K and $N_{CO}/\Delta V = 1.9\times 10^{16}$\,cm$^{-2}$\,km$^{-1}$\,s. The bullets display the results of these simulations for different volume density (n). We superimposed the R$_{21}$ line ratios computed in Table \ref{tab:depletion}, and R$_{31}$ ratio derived from the CO(3-2) JCMT observations of \citet{2018arXiv181210887L}. For the positions 15 and 28, there is a good velocity agreement, while \citet{2018arXiv181210887L} find a different CO(3-2) velocity for the position 36, when compared to our measurements. This is why there are 2 points corresponding to this position. The measurement uncertainties are displayed for each point. Interestingly, most R$_{21}$ points are incompatible with a critical density, as well as the R$_{32}$ points at $2\sigma$.}
\label{fig:lineratio} 
\end{figure}

The R$_{21}=$ CO(2-1)-to-CO(1-0) line ratio (in temperature) has been computed for each three detection and is provided in the last column of Table \ref{tab:depletion}. This ratio lies in the range 0.3-0.5, well below 1. Hence, we expect this gas to be sub-thermally excited. This is also a strong evidence that this gas is optically thick.
Furthermore, this quite low ratio supports the fact that we have not lost significant flux in the interferometric measurements in these regions detected both in CO(1-0) and CO(2-1). We expect that the detected gas is not extended but rather clumpy. One should note that the FWHM velocity width of these lines are in overall good agreement for CO(1-0) et CO(2-1), but for MC13-36. In addition, this clump has a velocity mismatch in \citet{2018arXiv181210887L}, so it should be considered with caution.

In parallel, there are clouds detected only in CO(2-1), that fall below our CO(1-0) sensitivity. There are 7 such clouds. 6 of them have a weak signal-to-noise ratio between 3\,$\sigma$ and 4\,$\sigma$. There is one cloud detected at 6\,$\sigma$ in CO(2-1) that is not seen in CO(1-0) at the offset (-9.0$^{\prime\prime}$,-8.3$^{\prime\prime}$). \citet{2018arXiv181210887L} has detected CO(3-2) molecular gas at this position and estimate a line ratio $R_{32} = 0.33\pm 0.11$.  Given the CO(1-0) $3\sigma$ upper limit (S$_{CO}^{1-0} < 1.14$ Jy\, km\, s$^{-1}$), we estimate a temperature ratio R$_{{2-1}/{1-0}} \ge 0.83 \,(3\sigma)$, which is not very likely according to our modelling, and the CO(1-0) molecular gas might have
been missed due to interferometric spatial filtering. Relying on the CO(2-1) and CO(3-2) single-dish detections, it is probable that this clump has properties similar to the three other clumps analysed below.

The molecular gas is not in thermal equilibrium with the dust. We considered a kinetic temperature in the range $7-15$\,K and assumed an excitation temperature $T_{ex} = 5$\,K. First, assuming this gas is optically thick, we derive a beam filling factor 
$\eta_{bf} = T_{obs}/(T_{ex}-T_{bf}) \times \Delta V / \delta v$, where $T_{obs}$ is the peak main beam temperature, T$_{bg}$ is the cosmic microwave background, $\Delta V$ is the FWHM width of the line and $\delta V$ the velocity resolution. We consider CO(2-1) measurements as they are not affected by beam filtering. We thus estimate a mean $\eta_{bf}^{5K}$ value of 5.2\,$\%$. We check that we find a similar value if we consider an excitation temperature of 30K (4.4\,$\%$).  This would correspond to clumps with an effective size $d = \sqrt \eta_{bf} D_{Beam} \simeq 10$\,pc. This value is compatible with the measurements performed in Table \ref{tab:AllDat2} and the maps displayed in Figure \ref{fig:vignettes}.
Second, we derive the mean column density of molecular hydrogen for the 3 positions studied in this section and found $\langle N_H \rangle = 3 \times 10^{20}$\,cm$^{-2}$, within the range found for the whole sample. We assumed a standard CO abundance $N_{CO} = 10^{-4} \times N_H$. We consider as input for the RADEX simulations \citep{2007A&A...468..627V} a CO column density of $N^{^\prime}_{CO}/{\Delta V} = N_{CO}/\eta^{5K}_{bf}/{\Delta V} = 1.9\times 10^{16}$ cm$^{-2}$ km$^{-1}$ s$^{+1}$ considering a mean ${\Delta V} = 30\,km/s^{-1}$.

We display in Figure \ref{fig:radex} the line ratios as function of the hydrogen density. We run RADEX simulations \citep{2007A&A...468..627V} for three different kinetic temperatures. The densities corresponding to the measured line ratios are weak and imply low excitation temperatures in the range 4-9\,K. We find densities that are in the range 60-650\,cm$^{-3}$ (resp. 250-1600\,cm$^{-3}$) for T$_K =$ 15\,K (resp. 7.5\,K). We can further check the consistency of these density estimates. Considering one single effective clump with a mean mass of 8700\,M$_\odot$ and a molecular hydrogen density in the range 60-650\,cm$^{-3}$, we expect a typical size in the range 8-16\,pc. Alternatively, if we consider 5 clumps in the beam, the clumps will have a typical size of 5-10\,pc.

In Figure \ref{fig:lineratio}, we check that the ratio R$_{21}$ points towards low density values lower than critical for T$_K = 15$\,K. We then derive the R$_{31}$ ratio relying on the CO(3-2) measurements of \citet{2018arXiv181210887L}. Given the uncertainties, the R$_{31}$ ratios are compatible within $2\sigma$.

In summary, we detect gas in non-local thermal equilibrium, with a low excitation temperature ($T_{ex}\sim 5-9$\,K). The kinetic temperature around 15\,K (and possibly smaller)  is significantly weaker that the dust temperature estimated in the infrared around 30\,K. The volume density of molecular gas is well below the critical density.

\begin{figure*}
\centering
\includegraphics[width=0.96\textwidth,clip]{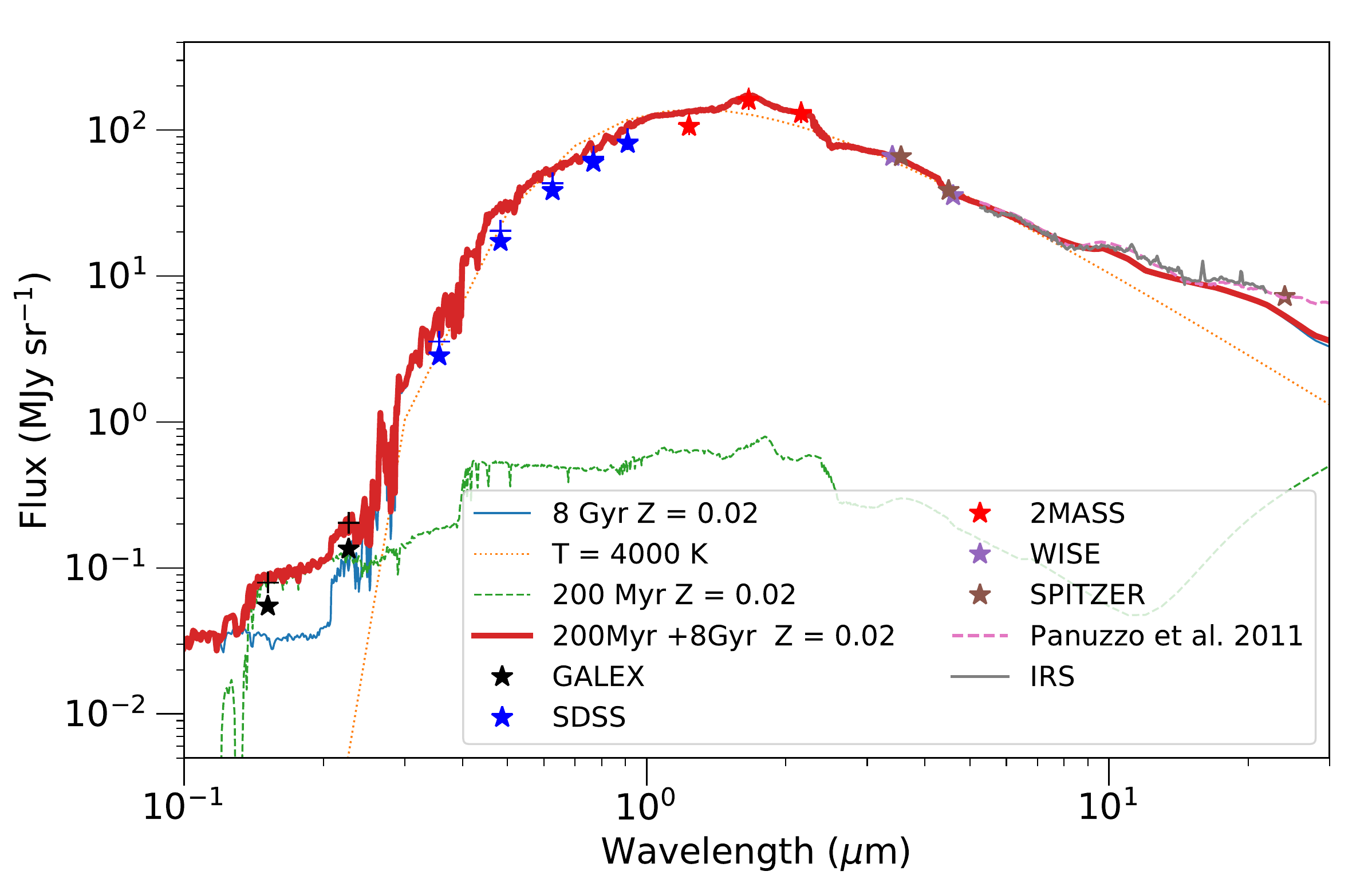} 
\caption{Spectral energy distribution integrated over the region observed with IRAM-PdB, excluding the region with the UV Bright nuclear disc \citep{2012ApJ...745..121L}. As described in Sect. \ref{sssect:SED}, we integrate over this region each photometric map available in this region after being reprojected on the 24$\mu$m Spitzer map and convolved to a 7\,arcsec FWHM Gaussian \citep{2011PASP..123.1218A}.  The star symbols correspond to the values thus achieved for each survey, while the crosses display the same measurements after a Galactic extinction correction (E(B-V)$=$0.05). In grey, we display the integrated IRS spectra from \citet{Hemachandra2015}, and the passive template from \citet{2011A&A...528A..10P}. Last, we superimpose different stellar templates from \citet{1998A&A...332..135B} and \citet{1999astro.ph.12179F}, as well as a 4000\,K black body. We show that GALEX points support solar metallicity templates, with a small contribution of a young stellar population (200\,Myr)}.
\label{fig:HemaSpect}
\end{figure*}

\subsection{Tracing star formation}
\label{ssect:TSF}
While the central region of Andromeda does not reveal any obvious star forming region, global star formation estimates usually display a non-zero star formation rate (SFR) in this region known to host ionised gas and dust. However, depending on the star formation tracers, quite different values have been estimated, even though this does not affect global SFR estimates \citep{2014A&A...567A..71V,2013ApJ...769...55F}, while the bulge is also often avoided \citep[e.g.][]{2015ApJ...805..183L,2017ApJ...834...70L}.

In distant galaxies, it is customary to  estimate the amount of obscured star formation observed in FUV by correcting the observed emission from young stars with the observed 24\,$\mu$m Spitzer flux. This has been studied statistically by \citet{2008AJ....136.2782L} and on sub-kpc scale by \citet{2008AJ....136.2846B}, and applied to Andromeda by \citet{2013ApJ...769...55F}. The FUV maps are expected to trace mainly the O and early B stars. It can be severely obscured by dust, while the UV emission from young stars is expected to heat and re-radiate in the mid-IR. The idea is hence to remove the contribution from the stellar emission and to perform a linear combination of the FUV and 24\,$\mu$m fluxes calibrated as proposed originally by \citet{2007ApJS..173..267S} and further tested by \citet{2008AJ....136.2782L}.

The region studied here is about 1/10$^{th}$ of the region studied as the bulge in \citet{2013ApJ...769...55F}. It is dominated by the stellar light with a very strong gradient. In this section, we reinvestigate this estimation of the SFR in the region where we observed molecular gas. In Sect. \ref{sssect:SED}, we gather and reproject a series of publicly available FUV-to-24$\mu$m images to study the spectral energy distribution of this region. In Sect. \ref{sssect:templ}, we compare the SED with various templates. On this basis, we show that an estimate of the SFR based on FUV and 24$\mu$m emissions corrected from stellar emission detect features at the noise level.

\subsubsection{Spectral energy distribution}
\label{sssect:SED}
We gather photometric maps obtained in the past ten years on the Andromeda galaxy. While we are interested in the central kilo-parsec field, it reveals important to get a full mosaic of the galaxy as the background determination is challenging and residuals prevent any photometric use of the maps. 
In the UV, we used the FUV and NUV GALEX Atlas of Nearby Galaxies \citep{2007ApJS..173..185G}. We retrieve 1\,square degree ugriz SDSS mosaic from Image Mosaic Service (Montage) at IPAC. We get the JHK 2MASS Large Galaxy Atlas \citep{2003AJ....125..525J}. We retrieve IRAC 3.6$\mu$m and 4.5$\mu$m and MIPS 24$\mu$m maps from Spitzer Heritage Archives. IRAC 5.8$\mu$m and 7.9$\mu$m maps suffer from bad background estimates in the central kilo-parsec and we were not able to use them\footnote{We also have to remove a constant background of 2.5\,MJy\,sr$^{-1}$ in order to get results consistent with \citet{2013ApJ...769...55F}.}. Last, we used the WISE image services to get the 3.4$\mu$m and 4.6$\mu$m maps found in excellent agreement with the IRAC Spitzer maps obtained at similar wavelengths. 

\begin{figure*}
\centering
\includegraphics[width=0.96\textwidth,clip]{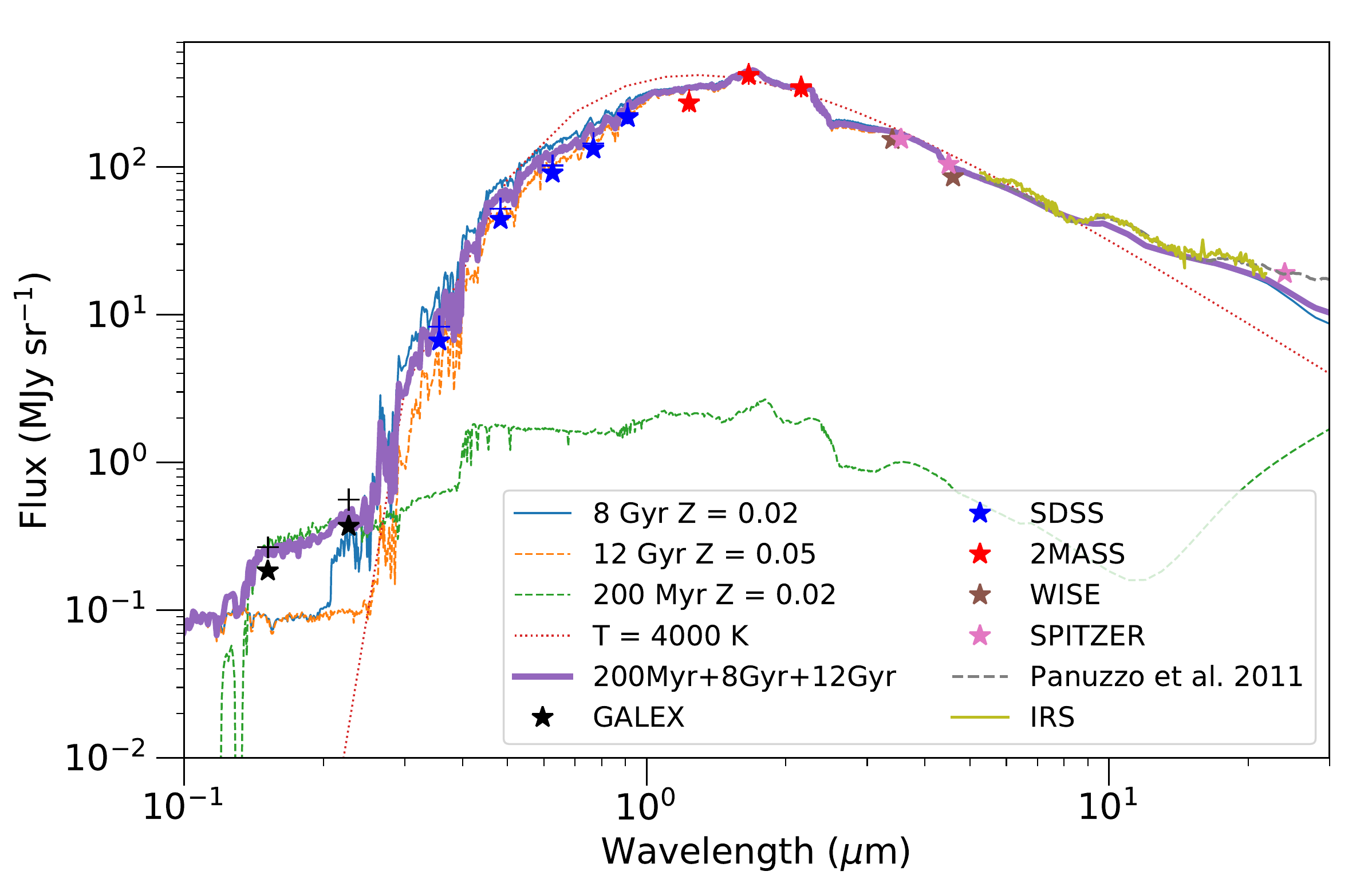} 
\caption{Spectral energy distribution integrated over the nuclear disc containing an FUV bright excess (with an FUV flux non corrected for Galactic extinction larger than 0.15\,MJy\,sr$^{-1}$\,kpc$^{-2}$). The same data sets and template library as those displayed in Figure \ref{fig:HemaSpect} are used.  In this case, GALEX points are more difficult to reproduce: beside the need of a larger contribution 200\,Myr stellar population, a fraction of a 12\,Gyr large metallicity template et required in addition to the 8\,Gyr solar metallicity template.}
\label{fig:HemaSpectUVB}
\end{figure*}
\begin{figure}
\centering
\includegraphics[width=0.48\textwidth,clip]{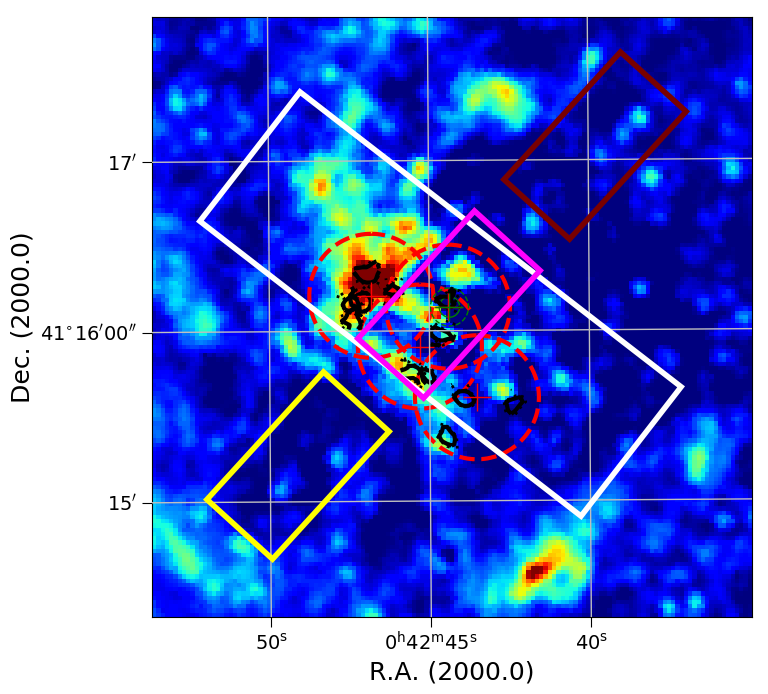} 
\caption{Map of $\Sigma_\text{SFR}$ from \citet{2013ApJ...769...55F} on which we show the regions observed with \textit{Spitzer}/IRS by \citet{Hemachandra2015} with rectangles. Each rectangle corresponds to a region observed by a different module of \textit{Spitzer}/IRS. The white rectangle correspond to the LL2 map, the red is SL2 and the yellow is SL1 and the magenta is an overlapped region between both SL1 and SL2. The dashed red circles are the fields of our data cube and the black contours correspond to the selected clumps.}
\label{fig:HemachandraRegions}
\end{figure}

In order to study the spectral energy distribution on a pixel basis, we first convolve each map to a 7\,arcsec FWHM Gaussian relying on the convolution technique based on kernels proposed by \citet{2011PASP..123.1218A}. Second, we reproject all the maps on a 2.4\,arcsec grid corresponding to 24$\mu$m Spitzer maps with Kapteyn Package. We can then plot the spectral energy distribution for each pixel and/or in an integrated area. To compute the spectral energy distribution displayed in Figures \ref{fig:HemaSpect} and \ref{fig:HemaSpectUVB}, we reproject the maps on the nuclear region observed by \citet{Hemachandra2015} with \textit{Spitzer}/Infrared Spectrograph (IRS). The nucleus was observed with the IRS short-low (SL1, SL2 and SL3) and long-low (LL2) modules which allows observations over the 5.2--20.75 $\mu$m band. The maps sizes were obtained by 18 overlapping observations for SL (32$^{\prime\prime}\times$57$^{\prime\prime}$) and 11 overlapping observations for LL (58$^{\prime\prime}\times$168$^{\prime\prime}$), both regions overlapping in the central area (see Figure~\ref{fig:HemachandraRegions}). The spectrum was integrated over the central region from which we cropped the sides since they contained clear border effects, including strong negative signals. SL2 (5.2--7.6 $\mu$m) and SL1 (7.5--14.5 $\mu$m) were connected using the overlapping SL3 (7.33--8.66 $\mu$m) mean flux. SL3 showed no offset with SL2, so the offset was added to SL1. SL1 was connected to LL2 (14.5--20.75 $\mu$m) with an offset to LL2 in order to obtain a coherent continuity with \textit{Spitzer} IRAC (4.5 $\mu$m) and MIPS (24 $\mu$m) measurements.

Figure \ref{fig:HemaSpect} corresponds to the field of view observed with IRAM-PdB, but excluding the FUV bright region (with a FUV flux non-corrected for Galactic extinction smaller than 0.15\,MJy\,sr$^{-1}$\,kpc$^{-2}$). We find a good agreement with an 8\,Gyr stellar templates from  \citet{1998A&A...332..135B} and a small contribution from a 200\,Myr stellar population (PEGASE.2, \citet{1999astro.ph.12179F}). The crosses included a correction of foreground extinction \citep{1999PASP..111...63F} assuming $E(B-V)=0.05$ following \citet[e.g.][]{2012ApJS..200...18D} and $R_V=3.1$.  This is in general agreement with the current belief that this region is dominated by old stellar population \citep[e.g.][]{2012ApJ...755..131R}. 
In the near infrared, we observe an excellent agreement with all stellar templates below 5\,$\mu$m.   We do find an excellent agreement of the near infrared measurements with the \citet{Hemachandra2015} IRS Spitzer spectra, with a significant silicate 'bump' around 10\,$\mu$m . It is also very close to the passive galaxy template displayed in Figure 2 of \citet{2011A&A...528A..10P}. This is typical of early-type galaxies (ETG) as discussed by \citet{2011A&A...528A..10P} and \citet{Rampazzo_2013} and more generally old stellar populations.  While \citet{2010A&A...509A..61S} observe a metallicity gradient with slit spectroscopy but with a large uncertainty, metal-rich $Z=0.05$, $\le$ 10-Gyr templates from \citet{1998A&A...332..135B} exhibit an infrared excess incompatible with the 24$\mu$m Spitzer data. Solar metallicity SSP templates of 8\,Gyr and 200Myr adjust correctly the data. Large metallicity templates improve the match with the data in the optical part but not in the UV nor IR (for an age smaller or equal to 10\,Gyr). One can note the presence of weak fine-structure emission lines discussed in \citet{Hemachandra2015}, which confirms the presence some interstellar ionised gas. One can point to relatively large [NeIII] emission line, often correlated to the 24\,$\mu$m dust emission \citep[e.g.][]{2013ApJ...777..156I}.

Figure \ref{fig:HemaSpectUVB} restricts to the FUV bright region excluded in Figure \ref{fig:HemaSpect}. This region is known to host an UV bright 200\,Myr stellar cluster as studied by \citet{2012ApJ...745..121L}. It is more difficult to match the data with a single template, suggesting a mixture of stellar population from different origins. The contribution of the 200\,Myr is more important, and it is also necessary to add a large metallicity template to reproduce the UV and optical part of the spectral energy distribution.  Here, the stellar background is stronger and the fine-structure emission lines are relatively weaker. Again, the FUV emission can be accounted for by stellar templates. There is no sign of star formation, while some atomic ionised gas and weak dust emission (see below) are tracing the interstellar medium of this region,  probably heated by the stellar population \citep{2012MNRAS.426..892G}.

\subsubsection{Dust emission and star formation estimate}
\label{sssect:templ}
\begin{figure*}
\centering
\includegraphics[width=0.96\textwidth,clip]{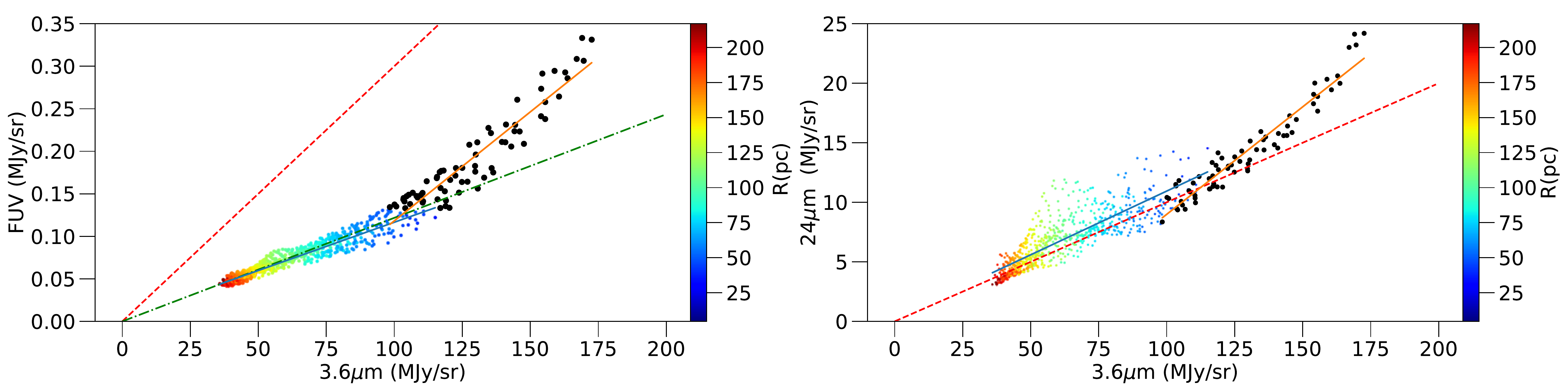} 
\caption{Correlations of the FUV and 24$\mu$m flux with the 3.6$\mu$m flux in the field of view observed with IRAM-PdB. The colour coding corresponds to the radial distance to the centre. The black large points correspond to measurements with a FUV flux (non corrected for Galactic extinction) larger than 0.09\,MJyr\,sr$^{-1}$. The red dashed lines correspond to the correction used by \citet{Leroy2013}, the green dashed line the correction used by \citet{2013ApJ...769...55F}. The blue and orange lines correspond to the linear adjustments performed on the two regions and tentatively used to correct the FUV and 24$\mu$m fluxes from stellar population.}
\label{fig:FluxCorr}
\end{figure*}
\begin{figure*}
\centering
\includegraphics[width=0.96\textwidth,clip]{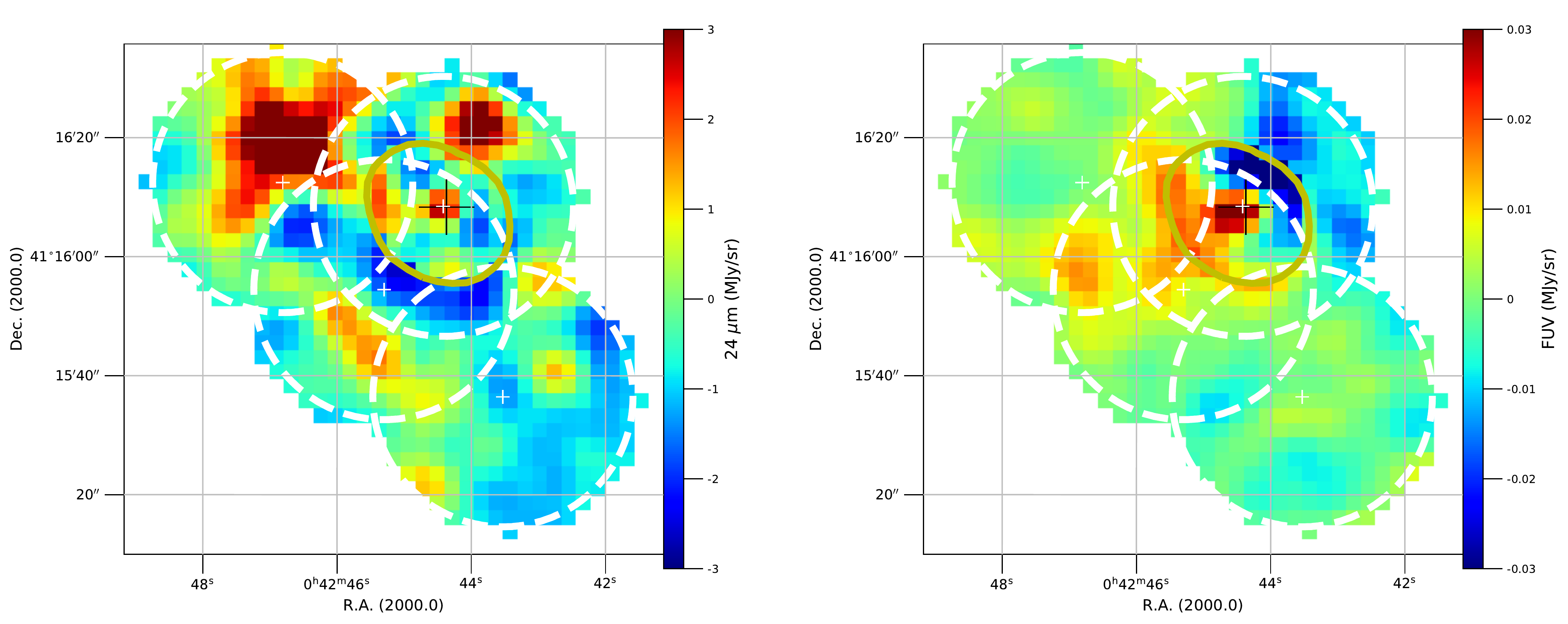} 
\caption{Different components used in star formation estimates. Following \citet{2008AJ....136.2782L} and \citet{2013ApJ...769...55F}, we subtract the stellar component on the 24$\mu$m Spitzer map (left panel) and on FUV GALEX map (right panel). The positions of the IRAM-PdB interferometer CO(1-0) observations are displayed with white dashed circles. The yellow contour corresponds to the FUV flux (non corrected for Galactic extinction) of 0.09\,MJyr\,sr$^{-1}$.}
\label{fig:SFRsuper}
\end{figure*}
\begin{figure*}
\centering
\includegraphics[width=0.96\textwidth,clip]{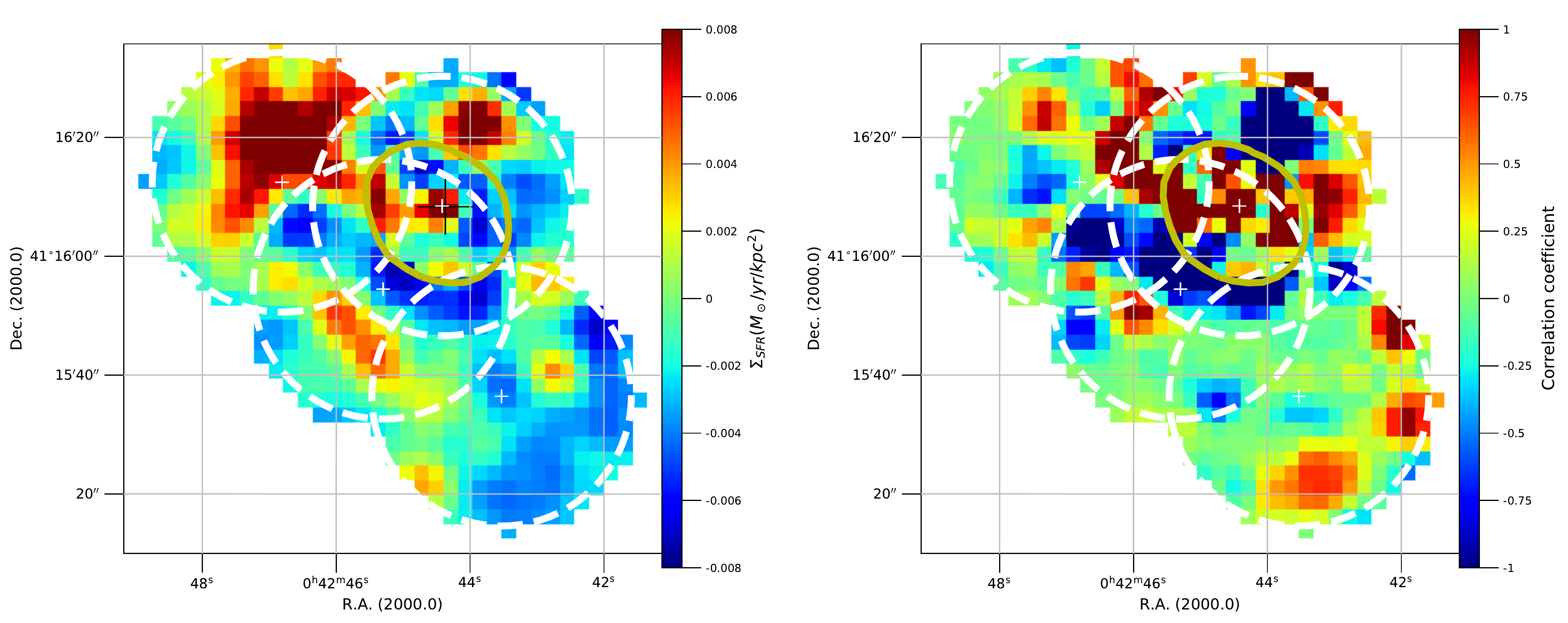} 
\caption{Star formation estimate derived the FUV and 24$\mu$m components displayed in Figure \ref{fig:SFRsuper} (left panel), and the correlation factor between these two components (right panel). The positions of the IRAM-PdB interferometer CO(1-0) observations are displayed with white dashed circles. The yellow contour corresponds to a FUV GALEX flux (non-corrected for Galactic extinction of 0.09\,MJy/sr).}
\label{fig:SFRmap}
\end{figure*}

\citet{Leroy2013} used a SFR estimate based on FUV and 24\,$\mu$m. This method has been used by \citet{2013ApJ...769...55F} who seem to overestimate the SFR in the circum-nuclear region with respect to \citet[e.g.][]{2014A&A...567A..71V}. In Figure \ref{fig:FluxCorr}, we compute the correlation between the FUV and 24\,$\mu$m with 3.6\,$\mu$m as computed by \citet{2013ApJ...769...55F}, with a colour-coding corresponding to the projected distance to the centre. 
These authors used the 3.6\,$\mu$m as an estimate of the stellar contribution to the 24\,$\mu$m and FUV for a 3.6\,$\mu$m flux smaller than 100\,MJy\,sr$^{-1}$.  We can note that there is a non-linear increase for a 3.6\,$\mu$m flux above 100\,MJy\,sr$^{-1}$ observed both in FUV and at 24\,$\mu$m. According to the spectral energy distribution of the region displayed in Figure \ref{fig:HemaSpectUVB}, this can be accounted for by a mixture of stellar populations. 

Figure \ref{fig:SFRsuper} displays the 24\,$\mu$m and FUV emissions after the subtraction of the emission from the stellar population performed with the adjustments displayed in Figure \ref{fig:FluxCorr}. We do detect in the 24\,$\mu$m map dust emission in the East side of the IRAM-PdB observations corresponding to three high SNR$_\textrm{tot}$ ($>45$) CO clumps, as displayed in Figure \ref{fig:CloudMapsVelo}, and corresponding to the MC13 $\#$28 and $\#$36 positions, CO(2-1) has been detected. We can note that weak 11.2\,$\mu$m is detected in this region observed at the edge of the IRS/Spitzer field \citep{Hemachandra2015}. In addition, a 24\,$\mu$m clump is observed 15$^{\prime\prime}$ North of the nucleus, where weak 11.2\,$\mu$m  and 17\,$\mu$m PAH features are detected.  No molecular gas has been detected there.
In these positions, no 6-9$\mu$m PAH features escape detection even though \citep{Hemachandra2015} discuss the importance of template subtractions. However, such 6-9 $\mu$m to 11.2$\mu$m PAH contrast is often observed in elliptical galaxies \citep{2005ApJ...632L..83K,2010ApJ...721.1090V}. 
Beside a very weak feature next to the nucleus, the other 24\,$\mu$m features are at the level of noise. In FUV, there is a bright spot next to the centre, which most probably corresponds to the subtraction of the FUV stellar cluster, while there might be some extinction on the North-West side of the field, corresponding to the inclination of the main disc. These features correspond to about 10\,$\%$ of the stellar continuum in this region and their amplitude is very sensitive to the linear correction adopted (see Figure \ref{fig:FluxCorr}).
When we combined these 2 maps following \citet{2008AJ....136.2782L}, we find the SFR map on the left hand side of Figure \ref{fig:SFRmap}. It seems dominated by the dust emission discussed above, that is most probably heated by the interstellar radiation field \citep{2012MNRAS.426..892G}. While this map averages to zero with a standard deviation of 0.004\,M$_\odot$\,yr$^{-1}$, the correlation between the FUV and 24\,$\mu$m maps displayed in Figure \ref{fig:SFRsuper} is shown in the right panel of Figure \ref{fig:SFRmap}. \citet{2008AJ....136.2782L} discussed that in case of star formation, a correlation is expected between the 24\,$\mu$m dust emission and the FUV extinction. We can observe here that we do observe a correlation in the North-West side, corresponding to the PAH dust emission discussed above, which supports the presence of dust with no molecular gas detection. There is also a correlation in the middle of the field, but it corresponds to a FUV-excess and negative dust feature. It is possibly an artefact at the limit (0.09MJy/sr in FUV flux) used to correct the stellar contribution, and does not correspond to an effect of star forming region.  The regions where we detect molecular gas do not exhibit a correlation coefficient close to $-1$, as can be seen with a comparison of Figure \ref{fig:CloudMapsVelo} with the right panel of Figure \ref{fig:SFRmap}. This supports the view that there is residual gas in this region that is not forming star, and some dust is heated by the intermediate and old stellar population.

\subsection{Kennicutt-Schmidt law}
\label{ssect:KSLaw}

\begin{figure}[h!]
\centering
\includegraphics[width=0.48\textwidth,clip]{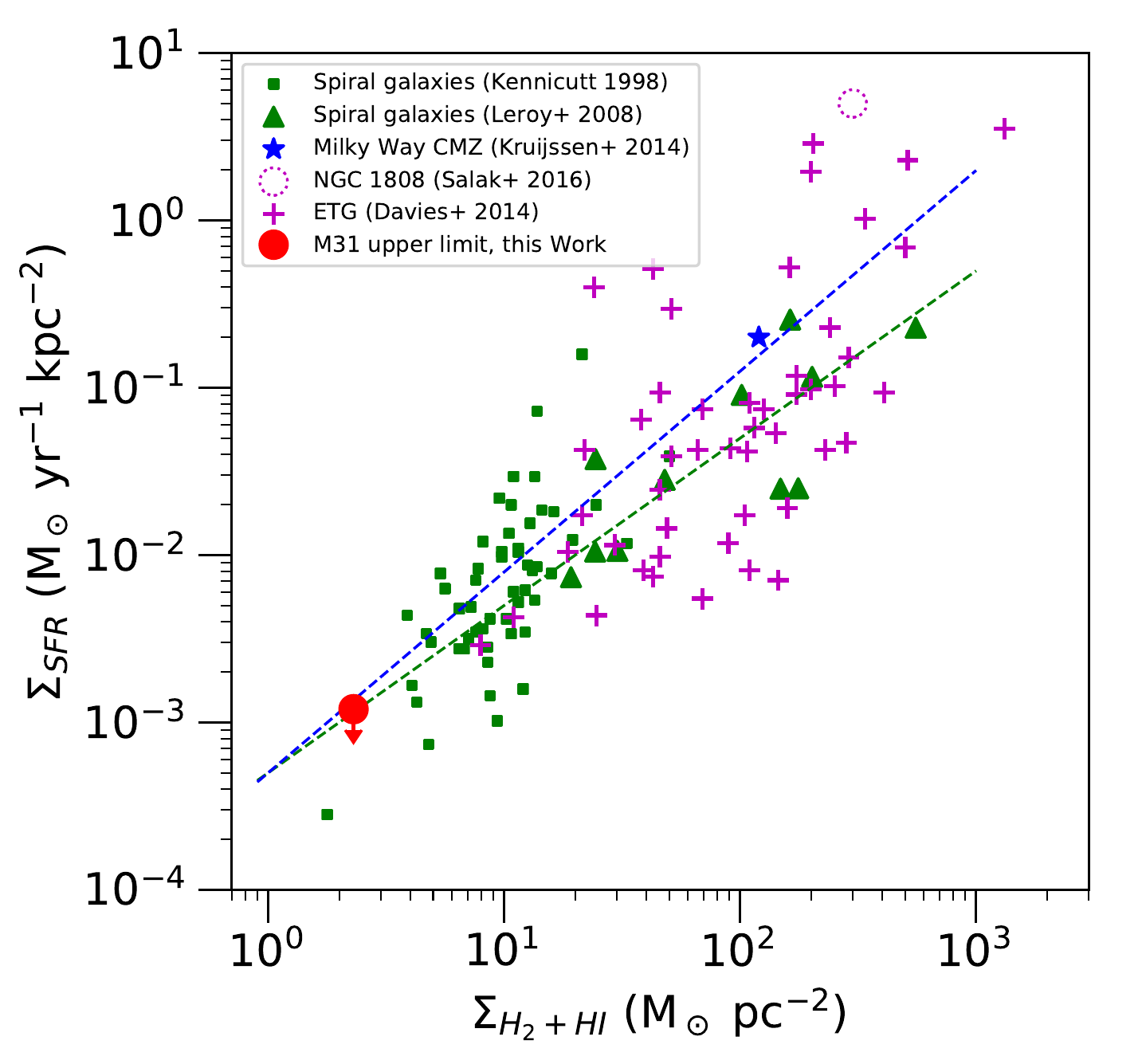}      
\caption{Surface density of SFR with respect to surface density of H$_2$+HI gas for various galaxies. The red filled circle is for M31 circum-nuclear region, where $\Sigma_{\textrm{H}_2+\textrm{HI}}$ has been calculated from our molecular gas measurements and the estimated surface density of HI in the corresponding region of M31 from \citet{Braun2009} (between 1 and 2 M$_{\odot} \, pc^2$) and $\Sigma_\textrm{SFR}$ from our FUV+24 $\mu$m our map displayed in the left panel of Figure~\ref{fig:SFRmap}. The green triangles are for the central region of spiral galaxies from \citet{2008AJ....136.2782L}. The magenta dotted circle is the circum-nuclear region (r < 200 pc) of NGC 1808 with $\Sigma_{\textrm{H}_2}$ from \citet{Salak2016} and $\Sigma_\textrm{SFR}$ estimated from \citet{Krabbe1994}. The magenta crosses are the central parts of early-type galaxies from \citet{Davis2014}. The green squares are spiral galaxies from \citet{Kennicutt1998}. The green square close to M31 is NGC~4698, an early-type spiral with a ring-like disk of star formation, and a decoupled nuclear disk \citep{Corsini2012}. The green dashed line is the Kennicutt-Schmidt law with a slope N = 1.0, and the blue dashed line is the linear regression applied on the whole sample.}
\label{fig:KS}
\end{figure}
M31 quiescence in its central region is made strikingly clear in our measurements. We estimate in the previous section an upper limit on the surface density SFR $\Sigma_\mathrm{SFR} < 0.0012 (3\sigma)$\,M$_\odot$\,yr$^{-1}$\,kpc$^{-2}$. In Table \ref{tab:AllDat2}, we estimate a mean molecular gas surface density of the detected clumps $\Sigma_\mathrm{H_2} =15$ \,M$_\odot$\,pc$^{-2}$, but when averaged on the region (of equivalent radius of 165\,pc), this density reduces to  $\Sigma_\mathrm{H_2} =0.9$\,M$_\odot$\,pc$^{-2}$.  This is a lower limit as smaller clumps have escaped detection. Last, we rely on \citet{Braun2009} to estimate a neutral hydrogen surface density of the region of $\Sigma_\mathrm{HI} =2 \pm 0.5$\,M$_\odot$\,pc$^{-2}$. In Figure~\ref{fig:KS}, we compare to the Kennicutt-Schmidt law (N = 1.4) with data from \citet{Williams2018, 2013ApJ...769...55F, Bigiel2010} and \cite{2008AJ....136.2782L}. Our upper limit is compatible with surface density SFR derived from \citet{2013ApJ...769...55F}, \citet{2014A&A...567A..71V}.

The circumnuclear region of Andromeda is typically not forming stars and we reach the limits as discussed by \citet{2012ApJ...752...98C}. One can argue that the region considered is small but we nevertheless integrate along the line of sight given the inclination of the disc.

Following \citet{2008AJ....136.2782L}, we define Star Formation Efficiency (SFE) as the SFR surface density per unit H$_2$, SFE(H$_2$) = $\Sigma_\textrm{SFR}/\Sigma_{\textrm{H}_2}$. Hence, we find a star formation efficiency of H$_2$ smaller than 1.3$\times$ 10$^{-9}$ yr$^{-1}$ (3$\sigma$), which is within the uncertainties of \citet{2008AJ....136.2782L} who finds SFE(H$_2$) = 5.25 $\pm$ 2.5 $\times$ 10$^{-10}$ yr$^{-1}$ at scales of 800 $pc$. A constant SFE(H$_2$) is expected if the GMC properties are universal and the conditions within the GMC constitute the driving factor \citep{Krumholz2005}. 

\section{Conclusion}
\label{sect:concl}
We have analysed molecular observations made with the IRAM-Plateau de Bure interferometer,
towards the central 250\,pc of M31. The CO(1-0) mosaic was compared with previous CO(2-1)
single dish mapping with the IRAM-30m HERA instrument. The spectral energy distribution of the region and tracers of star formation have been studied. 
  The main results can be summarized as follow :
\begin{enumerate}
\item We first identified molecular clumps from high S/N peaks in the map,
and selected genuine clouds using both robust data analysis and statistical methods. 
Our modified approach to the CPROPS algorithm  allowed us to extract a 
catalogue of 12 CO(1-0) clumps with a total S/N ratio larger than 15 in a region of low CO density.
\item We derived the size, velocity dispersion, H$_2$ mass, surface density and virial parameter for each cloud. The clouds follow the Larson's mass-size scaling relation, but lie above the velocity-size relation. We discuss that they are not virialised, but might be the superimposition of smaller entities.
\item We measured a gas-to-stellar mass ratio of $4\times 10^{-4}$ in this region.
\item We discussed the CO(2-1)-to-CO(1-0) line ratio from the single dish observations from \cite{2013A&A...549A..27M}  when the CO(1-0) map is smoothed to the 12\,arcsec resolution. We have been able to compute the CO(2-1)-to-CO(1-0) line ratio for 3 positions. This ratio is below 0.5, supporting non LTE conditions. Relying on RADEX simulations, we discuss that this optically thick gas has a low gas density in the range $60-650$\,cm$^{-3}$ with sub-thermal excitation ($T_{ex} = 5-9$\,K. We find a filling factor of order 5\,$\%$.
\item We study the spectral energy distribution of the region, and show that it is compatible with a quiescent elliptical galaxy template, with weak MIR atomic fine-structure emission lines. These lines correlates with the 24$\mu$m dust emission detected in the region.
\item A correlation between the FUV extinction and 24$\mu$m dust emission is observed in a position at 15\,arcsec North of the nucleus, where \citet{Hemachandra2015} have detected 11.2 and 17.\,$\mu$m PAH. No molecular gas is detected in this position. In the region located East of the nucleus, where several molecular clouds have been detected, there is 24$\mu$m dust emission as well as 11.2\,$\mu$m PAH emission \citep{Hemachandra2015}.
\item We subtracted the stellar contribution in order to use 24\,$\mu$m and FUV as tracers of star formation. This region averages to zero star formation, deriving an upper limit compatible with previous works. This low density region (both in gas and SFR) lies formally on the  Kennicutt-Schmidt law.
\end{enumerate}

The circum-nuclear region of M31 appears depleted in gas, both HI and H$_2$. This depletion is compatible with the absence of star formation. The gas and dust observed in this region are heated in the centre by the radiation of intermediate and old stellar populations. The galaxy
then appears to be quenched from inside out, since star formation continues to occur in 
the outer ring. The presence of diffuse ionized gas in the centre, together with 
low abundance of neutral gas (atomic and molecular) and lack of young stars in the centre is quite similar to the LIER galaxies described by \citet{2016MNRAS.461.3111B}. 

In the case of M31, many scenarios have been proposed. 
\citet{Dong2018} observations suggest that there was a strong peak in SFR less than 500 Myr ago, 
which could contribute to the gas depletion. This is in line with the conclusions of \citet{2006Natur.443..832B} which 
point at a collision with M32 as a reason for the low molecular gas density in the centre of Andromeda, 
also explaining the atypical two-rings architecture of M31. 
\cite{DSouza2018} claim that M32 was in the past a much bigger galaxy, able to produce a major
interaction with M31 2 Gyr ago. This allows also the remaining M32-core to re-enter the M31 disk almost 
head-on towards the centre and produce more recently the two rings in the disk. 
  Observations of higher-J CO emission, and also shock-tracing molecules will be
helpful to disentangle the various scenarios.

\begin{acknowledgements}
Based on observations carried out with the IRAM Plateau de Bure Interferometer. IRAM is supported by INSU/CNRS (France), MPG (Germany) and IGN (Spain).
We acknowledge the IRAM Plateau de Bure team for the observations. 
We thank Sabine K{\"o}nig for her support for the data reduction. We warmly thank Eric Rosolowsky, our referee, for his very helpful and constructive comments. We are grateful to Pauline Barmby for providing us with reduced Spitzer data in the nucleus region.
ALM has benefited from the support of \textit{Action Fédératrice Structuration de l'Univers et Cosmologie} from Paris Observatory.
This research has made use of the NASA/\(\)IPAC Infrared Science Archive, which is operated by the Jet Propulsion Laboratory, California Institute of Technology, under contract with the National Aeronautics and Space Administration. This research made use of Montage, funded by the National Aeronautics and Space Administration's Earth Science Technology Office, Computational Technnologies Project, under Cooperative Agreement Number NCC5-626 between NASA and the California Institute of Technology. The code is maintained by the NASA/IPAC Infrared Science Archive.
This research makes use of GALEX, SDSS, 2MASS, Spitzer and WISE archive data of Andromeda.
\end{acknowledgements}


\appendix

\section{Upper limit on the continuum level}
\label{sect:cont}
\begin{figure}
\centering
\includegraphics[width=0.48\textwidth,clip]{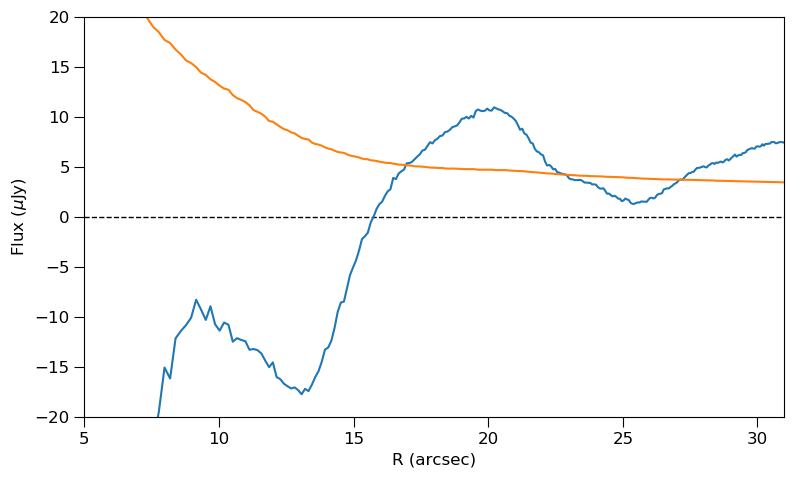}
\caption{Mean flux (blue line) and associated noise (orange line) in $\mu$Jy of the spectrum integrated within concentric circles of increasing radius R. Circles are centred on the centre of the field, the channels within the [-600, 0] km$\,$s$^{-1}$ range were removed to avoid influence of the detected clumps.}
\label{fig:continuum}
\end{figure}

We estimate an upper limit on the continuum level with an average on the whole bandwidth $[-3000,6000]$\,km\,s$^{-1}$. 

In Figure \ref{fig:continuum}, we display the mean flux and associated (rms) noise in $\mu$Jy achieved in integrating the whole spectra in circular regions of given radius R. 
We significantly improve the first 3$\sigma$ upper limit measured at 1~mm by Melchior $\&$ Combes (2013) of 0.65\,mJy with single dish observations. We estimate a 3\,$\sigma$ upper limit at 3~mm of about 15$\mu$Jy within a radius of 20 arcsec. 

This excludes the value (60 $\mu$Jy) at 12\,$\sigma$ expected with a simple extrapolation of the synchrotron emission detected at 6\,cm (5\,GHz)  of about 0.2 mJy by Giessubel $\&$ Beck (2014) with a measured spectral index of $-0.4\pm 0.03$. As the primary beam has a HBPW of 45\,arcsec, we do not expect to significantly smooth large scale signals. We thus expect a steeper slope.   

We can also exclude large quantities of hidden cold dust. However, it is already well-known that this region hosts very little dust \citep[e.g.][]{2000MNRAS.312L..29M}, and our upper limit does not provide more constraints than the infrared dust emission presented in the Figure 3 of Groves et al. (2012), as dust emission at 3~mm is expected at the level of 10$^{-6}$$\mu$Jy.  If a weak 3mm continuum exists in this region, it is probably due to synchrotron emission.

\section{Selection procedure}
\begin{figure}
\centering
\includegraphics[width=0.48\textwidth,clip]{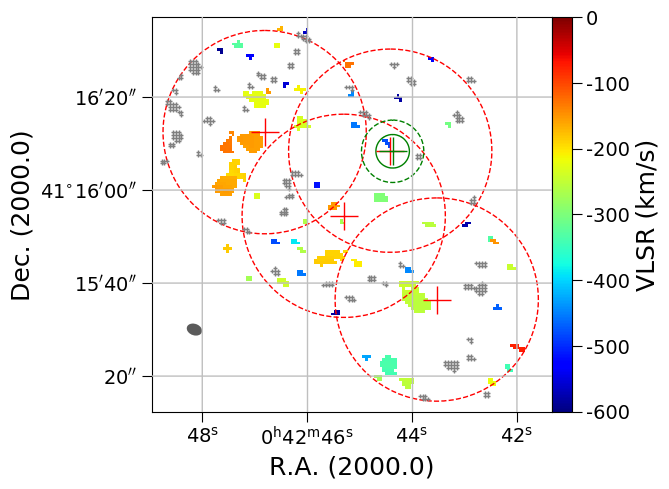}
\includegraphics[width=0.47\textwidth,clip]{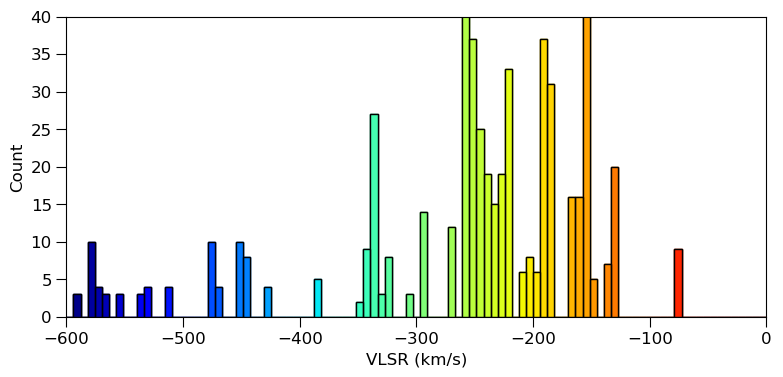}
\caption{Maps of the velocities core clumps detected (top) and associated velocity distribution (down). The colour coding for the velocity is the same for the maps and histograms. The velocity values for core clumps with contiguous pixels with more than 2 spectral channels above  3\,$\sigma_{noise}$ are displayed. Positions of the negative core clumps are displayed as grey crosses. The histograms of velocity reveal some spatial and spectral correspondences. Three core clumps with strong signal in particular can be observed at velocities -157, -187 and -253 km\,s$^{-1}$. This is used to apply \textsc{Gildas} cleaning algorithm on specific channels. Some cores are detected outside of the fields and a few also display abnormal velocities for the region studied (especially one at -76 km\,s$^{-1}$). All these clumps are rejected by our selection procedure.}
\label{fig:veloStudy}
\end{figure}
\begin{figure*}
\centering
\includegraphics[width=0.48\textwidth,clip]{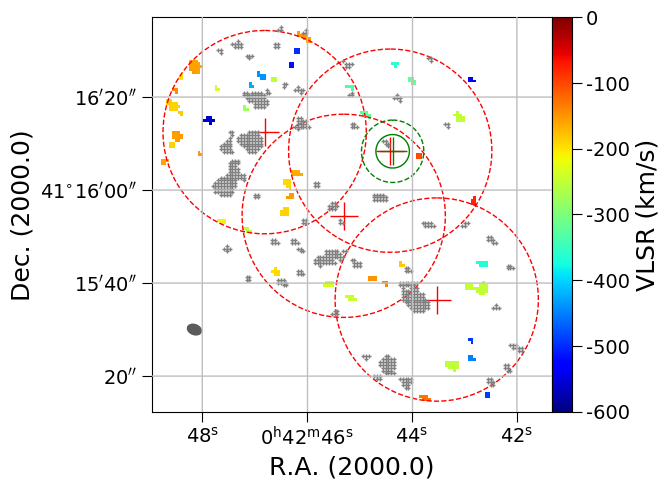}
\includegraphics[width=0.47\textwidth,clip]{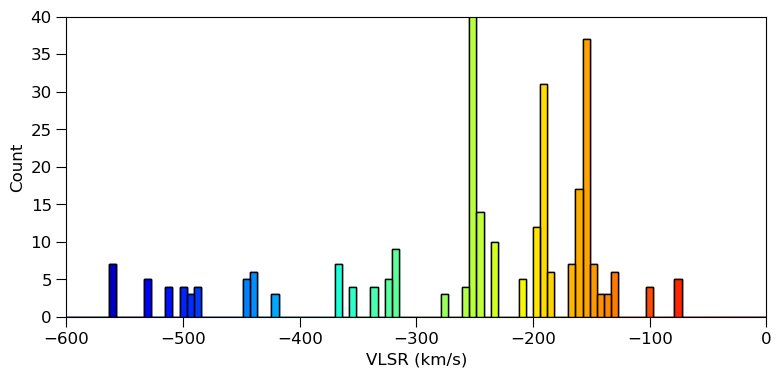}
\caption{Same as Figure \ref{fig:veloStudy} with the velocity values for core clumps with contiguous pixels with more than 2 spectral channels below 3\,$\sigma_{noise}$. Positions of the positive core clumps are displayed as grey crosses.}
\label{fig:veloStudy2}
\end{figure*}
\begin{figure*}
    \centering
    \includegraphics[width=0.96\textwidth,clip]{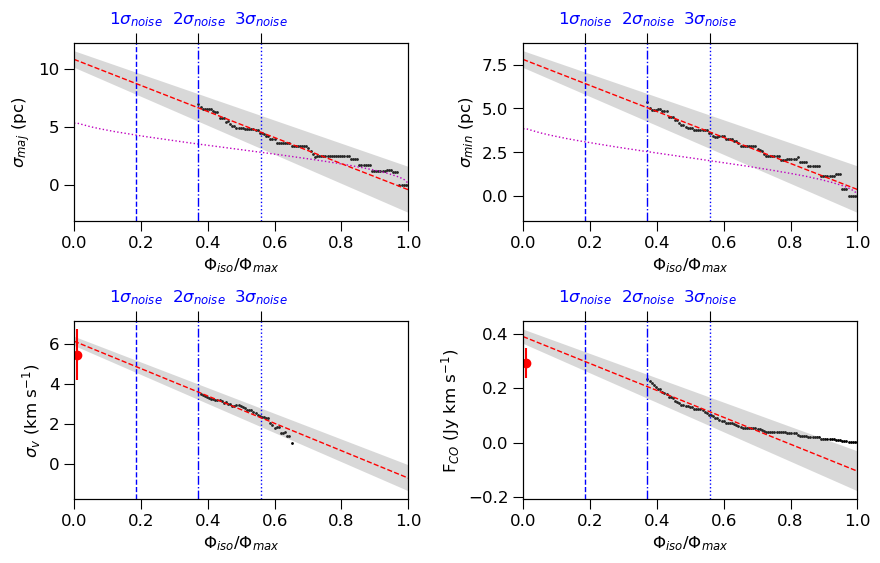}
    \caption{Same as Figure \ref{fig:Extrap1} for one clump at J2000 coordinates $00h\,42^m\,44.52^s$ and $+41^{\circ}\,16'\,10.1^{\prime\prime}$ with SNR$_{peak}=5.4$ and SNR$_{tot}=15.2$. Here, the size over the major and minor axis is twice as high as what is obtained from the simulated clump, which suggests this clump is resolved. The velocity dispersion is more than twice the spectral resolution (2.2\,km/s). There is also a good agreement between the extrapolated and measured through GILDAS velocity dispersions, while the total flux is slightly higher for the extrapolation.}
    \label{fig:Extrap2}
\end{figure*}

With a 2-D Gaussian model using the transfer function of a point-like signal we show that isolated pixels showing an excess are impossible to distinguish between signal and noise. We see that a minimum of 3 adjacent spatial pixels is required. We define as a cloud an area of at least 3 adjacent spatial pixels with signals with velocities $V_{0}$ varying within a range of $10\,km\,s^{-1}$ maximum. We find 54 molecular clouds (displayed in  Figure~\ref{fig:veloStudy}) with peak flux ranging from $\Phi_{max} = 11.8 \, mJy/beam$ to $\Phi_{max} = 60.1 \, mJy/beam$. $3-\sigma$ clumps with negative flux are also found, we display 47 negative clumps in Figure~\ref{fig:veloStudy2}. These negative signals are expected to be noise or side lobes: we keep them to test our selection procedure. The velocity distribution histograms for positive and negative clumps show that negative clumps with numerous pixels lie in the same regions as positive core clumps with high S/N ratio.

\end{document}